\documentclass[aps,prb,groupedaddress,showpacs,twocolumn,superscriptaddress,10pt]{revtex4-2}
\usepackage{graphicx}
\usepackage{mathtools}
\usepackage{amsmath}
\usepackage{amssymb}
\usepackage{bbm}
\usepackage{bm}
\usepackage{cancel}
\usepackage{xspace}
\usepackage{braket}
\usepackage{xcolor}
\usepackage{calrsfs}
\usepackage{comment}
\usepackage{placeins}%enables \FloatBarier
\usepackage[colorlinks=true]{hyperref} % Use hidelinks to hide highlighting

\usepackage{units}
 
\newcommand{\ee}{\mathrm{e}}  % Euler number
\DeclareMathOperator*{\ii}{i} % imaginary unit
\newcommand*\dd{\mathop{}\!\mathrm{d}}

\renewcommand{\vec}[1]{\bm{#1}} % vectors in bold
\newcommand{\mat}[1]{\bm{#1}} % matrices in bold
\newcommand{\kel}[1]{\underline{#1}} % objects on Keldysh contour

\definecolor{hblue}{RGB}{0,0,255}
\newcommand{\resub}[1]{{\color{black} #1}}

\begin{document}

\title{Correlated Mott insulators in strong electric fields: \\
Role of phonons in heat dissipation}
	
\author{T. M. Mazzocchi}
\email[]{mazzocchi@tugraz.at}
\affiliation{Institute of Theoretical and Computational Physics, Graz University of Technology, 8010 Graz, Austria}
\author{P. Gazzaneo}
\affiliation{Institute of Theoretical and Computational Physics, Graz University of Technology, 8010 Graz, Austria}
\author{J. Lotze}
\affiliation{Institute of Theoretical and Computational Physics, Graz University of Technology, 8010 Graz, Austria}
\author{E. Arrigoni}
\email[]{arrigoni@tugraz.at}
\affiliation{Institute of Theoretical and Computational Physics, Graz University of Technology, 8010 Graz, Austria}

\date{September 15, 2022}
	
\begin{abstract}

We study the spectral and transport properties of a Mott insulator driven by a static electric field into a non-equilibrium steady state. For the dissipation, we consider two mechanisms: Wide-band fermion reservoirs and phonons included within the Migdal approximation. The electron correlations are treated via non-equilibrium dynamical mean field theory with an impurity solver suitable for strong correlations. We find that dissipation via phonons is limited to restricted ranges of field values around Wannier-Stark resonances. To cover the full range of field strengths, we allow for a small coupling with fermionic baths, which stabilizes the solution. When considering both dissipation mechanisms, we find that phonons enhance the current for field strengths close to half of the gap while lowering it at the gap resonance as compared to the purely electronic dissipation used by Murakami and Werner \resub{[Murakami and Werner, Phys. Rev. B 98, 075102 (2018)]}. Once phonons are the {\em only} dissipation mechanism, the current in the metallic phase is almost one order of magnitude smaller than the typical values obtained by coupling to a fermionic bath. In this case, the transport regime is characterized by an accumulation of charge in the upper Hubbard band described by two effective chemical potentials.
						 
\end{abstract}
	
%71.10.-w Theories and models of many-electron systems
%71.10.Fd Lattice fermion models (Hubbard model, etc...)
%71.15.-m Methods of electronic structure calculations
%71.27+a Strongly correlated electron systems; heavy fermions
%72.15.Qm Scattering mechanisms and Kondo effect
%72.20.−i Conductivity phenomena and in semiconductors and insulators
%73.21.La 	Quantum dots (Electron states and collective excitations in multilayers, quantum wells, mesoscopic, and nanoscale systems)
%73.23.-b       Electronic transport in mesoscopic systems
%73.63.Kv       Quantum dots (Electronic transport in nanoscale materials and structures)
%77.22.Jp Dielectric breakdown and space-charge effects
	
\maketitle

\section{Introduction}\label{sec:intro}
One of the puzzling properties of strongly correlated materials is the occurrence of resistive switching in the form of an electric field-driven insulator-to-metal transition~\cite{ja.tr.15}. The physical mechanism(s) leading to such a process is, as of yet, debated. In particular, the two most established explanations rely on either a thermal-~\cite{li.ar.15,ha.li.18} or quantum-triggered~\cite{ha.ar.22u} origin of such phenomena.

Another possible explanation advocates the formation of conducting filaments percolating through the material. Recent work employed effective-resistor models~\cite{ja.tr.15,st.ca.13} or a non-homogeneous mean field approach~\cite{li.ar.17} to describe the formation of such filaments. 
It is beyond doubt that a comprehensive description of these phenomena requires a deeper understanding of the microscopic mechanism behind the Joule heat transport.

However, beyond the relevance of the resistive switching transition, the problem of non-equilibrium transport of correlated electrons in a strong dc electric field and the occurrence of current-carrying regimes has attracted considerable attention in recent years~\cite{zene.34,ok.ar.03,ok.ao.05,free.08,jo.fr.08,jo.fr.08,ec.ok.10,am.we.12,aron.12,ec.we.13,le.pa.14,ma.am.15,mu.we.18}. As already pointed out in previous works~\cite{ts.ok.09,ec.ok.10,am.we.12,aron.12,han.13,ec.we.13,ha.li.13,li.ar.17,mu.we.18}, in order for the system to reach a non-equilibrium steady-state, the current-induced Joule heat must be carried away. 
An efficient way to achieve this consists in the inclusion of fermionic reservoirs~\cite{ts.ok.09,aron.12,li.ar.15,mu.we.18,ha.li.18}, also known as \emph{B\"uttiker probes}. Their bandwidth is usually chosen to be large enough to ensure dissipation over the whole energy window of interest. \resub{Even though fermionic reservoirs proved themselves to be particularly well suited for the system to reach a stable, non-trivial steady-state solution~\cite{ar.ko.12,am.we.12}, they fail to capture the physical mechanism of heat transport mediated by lattice vibrations}. An important step in this direction can be found in~\cite{ha.ar.22u} where both phonons --- connected to a bosonic bath --- and fermionic reservoirs are included in two--dimensional Hubbard-like systems subject to a static field.

A description of the electric field-driven insulator-to-metal transition is beyond the purpose of the present work. We rather aim to model the heat dissipation of a Mott insulating system on a more realistic level by explicitly incorporating electron-phonon (e-ph) scattering as dissipation channel. Similarly to Ref.~\cite{mu.we.18}, we focus on the Hubbard model within the dynamical mean field theory (DMFT) approximation.
We find that the non-equilibrium picture near the current-carrying regions is significantly affected because of the different nature of the e-ph screening compared to the action of the fermionic baths. 

In particular, when only phonons contribute to dissipation, their relatively narrow band width makes it difficult to reach a stable steady-state~\footnote{\resub{For our purposes, a non-equilibrium steady-state is {\em unstable} if {\em either} the DMFT loop does not converge {\em or} convergence is achieved but the observables (current, double occupation, kinetic energy) display very rapid and large (in magnitude) oscillations as function of adjacent values of the applied electric field.}}  solution except for applied fields close to half of or the full band gap. Because of this limitation, addressing a wide region of field strengths requires coupling to a fermionic bath. We realize this mixed situation by weakly coupling to a fermionic B\"uttiker-like reservoir in addition to the phonons. Electronic correlations in this Hubbard model plus phonons are treated within DMFT using the auxiliary master equation approach (AMEA) impurity solver~\cite{ar.kn.13,do.nu.14,do.ga.15,ti.do.15,so.do.18}, which is particularly suited to treat strong correlations in a non-equilibrium steady-state.

The results obtained in the mixed situation agree with Ref.~\cite{mu.we.18} for both, the spectral features and physical quantities. With respect to the previous work, we also study the dependence of the observables on the coupling to the fermionic bath. We find that phonons enhance the steady-state current for field strengths of the order of half of the gap while having no effect at the gap resonance, provided that the coupling to the fermionic bath is sufficiently small. We also show that phonons alone are far less effective as heat reservoirs, thus suppressing the current, especially at the full gap resonance.

This paper is organized as follows: In Sec.~\ref{sec:MO_HA} we present the model Hamiltonian, while in Sec.~\ref{sec:method} we briefly describe Floquet-DMFT (F-DMFT)~\cite{ts.ok.08,sc.mo.02u,jo.fr.08} with the AMEA impurity solver. Our main results (Figs.~\ref{fig:PH_SFs_maxima} and \ref{fig:phonon_observables}) are discussed in Sec.~\ref{sec:results}, whereas Sec.~\ref{sec:conclusions} is left for final considerations. 
	
\section{Model Hamiltonian}\label{sec:MO_HA}
Similarly to Ref.~\cite{mu.we.18}, we consider an extension of the Hubbard model in the presence of an electric field described by the Hamiltonian 
\begin{align}\label{eq:MicroHamiltonian}
\hat{H}(t) &= \varepsilon_{\text{c}} \sum_{i\sigma}\hat{n}_{i\sigma} -\sum_{\sigma}\sum_{(i,j)} t_{ij}(t) \hat{c}^{\dagger}_{i\sigma} \hat{c}_{j\sigma} + U \sum_{i} \hat{n}_{i\uparrow} \hat{n}_{i\downarrow} \notag \\ 
&+ \hat{H}_{\text{bath}} + \hat{H}_{\text{e-ph}} + \hat{H}_{\text{ph}}.
\end{align}
The operator $\hat{c}^{\dagger}_{i\sigma}$ ($\hat{c}_{i\sigma}$) creates (annihilates) an electron of spin $\sigma= \{ \uparrow,\downarrow \}$ at the $i$-th lattice site, with the corresponding density operator $\hat{n}_{i\sigma}\equiv \hat{c}^{\dagger}_{i\sigma} \hat{c}_{i\sigma}$. Nearest neighbor sites are denoted by $(i,j)$ and the electrons' \emph{onsite energy} is $\varepsilon_{\text{c}} \equiv -U/2$. 
The static homogeneous electric field is introduced within the temporal gauge 
and enters in the time dependence of the hopping in Eq.~\eqref{eq:MicroHamiltonian} via the Peierls substitution~\cite{peie.33}
\begin{equation}\label{eq:peierls}
t_{ij}(t) = t_{\text{c}} \ \ee^{-\ii \frac{q}{\hbar} \left( \vec{r}_j - \vec{r}_i \right) \cdot \vec{A}(t)}, 
\end{equation} 
where $t_{\text{c}}$ is the lattice hopping amplitude, $\vec{A}$(t) is the vector potential, $q$ the electron charge and $\hbar$ Planck's constant. We choose the electric field $\vec{F}= -\partial_{t}\vec{A}(t)$ and thus $\vec{A}(t)=\vec{e}_{0} A(t)$ to point along the lattice body diagonal, i.e. $\vec{e}_{0}=(1,1,\ldots,1)$ and work in the temporal gauge
\begin{equation}\label{eq:TD_VecPot}
\vec{A}(t)= -\vec{F}\  t.
\end{equation}
Eqs.~\eqref{eq:peierls} and \eqref{eq:TD_VecPot} introduce the Bloch frequency  $\Omega \equiv -Fqa/\hbar$, with $a$ being the lattice spacing. 
In this work we consider a $d$-dimensional lattice in the $d \rightarrow \infty$ limit~\cite{mu.we.18} with the usual rescaling of the hopping $t_{\text{c}}=t^{\ast}/(2\sqrt{d})$. This way, sums over the crystal momentum $\sum_{\vec{k}} \chi(\omega,\vec{k})$ of generic quantities $\chi$ can be replaced by an integral~\cite{ts.ok.08}
$\int \dd\epsilon \int \dd\overline{\epsilon} \ \rho(\epsilon,\overline{\epsilon}) \chi(\omega;\epsilon,\overline{\epsilon})$, where $\rho(\epsilon,\overline{\epsilon}) = 1/(\pi t^{\ast 2}) \ \exp(-( \epsilon^{2} + \overline{\epsilon}^{2})/t^{\ast 2})$ is the joint density of states (JDOS) and
\begin{align}\label{eq:d-dim_crystal_dep}
\epsilon & = -2t_{\text{c}} \sum_{i=1}^{d} \cos(k_i a), \\
\overline{\epsilon}& = -2t_{\text{c}}\sum_{i=1}^{d} \sin(k_i a). \notag
\end{align}

Electron-phonon coupling is included in the form of an acoustic phonon branch attached to each lattice site
\begin{equation}\label{eq:e-ph_ham}
\hat{H}_{\text{e-ph}} = g \sum_{i} \left( \hat{n}_{i\uparrow} + \hat{n}_{i\downarrow} \right) \hat{x}_{i}, 
\end{equation}
where $\hat{x}_{i} \equiv (\hat{b}^{\dagger}_{i} + \hat{b}_i)/\sqrt{2}$ and  $\hat{b}^{\dagger}_{i}$ ($\hat{b}_{i}$) creates (annihilates) a phonon belonging to the branch $i$. We include the dispersion of each branch in $\hat{H}_{\text{ph}}$, which will be discussed in Sec.~\ref{sec:Dyson-eq}~\footnote{Phonon mixing between the branches is neglected in the spirit of the DMFT approximation and justified by the fact that we are only interested in the properties of the phonons as heat reservoirs.}.
 
As pointed out in the introduction, a stable steady-state is difficult to achieve when considering dissipation  by phonons only. Due to the narrow phonon dispersion, multiphonon processes would be necessary in order to relax electrons across the Hubbard gap. For this reason, we additionally include fermionic baths and try to extrapolate the properties of the system to the limit 
where the coupling to these baths become negligible. The fermionic baths are in the form of \emph{infinite} tight-binding (TB) chains attached to each lattice site~\cite{han.13,ha.li.13,ne.ar.15,li.ar.17}. The corresponding Hamiltonian reads
\begin{align}\label{eq:TBchains}
\hat{H}_{\text{bath}} &= - t_b \sum_{\alpha j \sigma} \left( \hat{f}^{\dagger}_{\alpha j \sigma} \hat{f}_{\alpha j+1 \sigma} + \text{h.c.} \right)+ \varepsilon_{b} \sum_{\alpha j \sigma} \hat{n}^{f}_{\alpha j \sigma} \notag \\
&\phantom{= }- v \sum_{j\sigma} \left( \hat{c}^{\dagger}_{j\sigma} \hat{f}_{1j\sigma} + \text{h.c.} \right),
\end{align}
where $\hat{f}^{\dagger}_{\alpha j \sigma}$ ($\hat{f}_{\alpha j \sigma}$) creates (annihilates) an electron with spin $\sigma$ at position $\alpha$ along the $j$-th bath chain and $\hat{n}^{f}_{\alpha j \sigma}$ is the corresponding density operator. The last term in Eq.~\eqref{eq:TBchains} accounts for the hybridization between the $j$-th site in the lattice and the corresponding bath chain. In this work we set $\hbar = k_{\text{B}} = a = 1 = -q$, such that $\Omega \equiv F$, i.e. current and electric field are measured in units of $t^{\ast}$. 

\section{Method and formalism}\label{sec:method}

\subsection{Floquet Green's functions}\label{sec:GFs_Dyson_Floquet}
Due to the periodicity of the non-equilibrium Green's functions in the temporal gauge~\footnote{It should be noted that in the case of the dc field considered here, the system is time-independent and so is its GF. Since we choose to work within the temporal gauge, we stick to the Floquet-formalism for periodically driven systems, thus keep using expressions like characteristic period and frequency.}, $G(t,t^{\prime})=G(t+\tau,t^{\prime}+\tau)$ with $\tau=2\pi/\Omega$, it is convenient to represent them in the \emph{Keldysh-Floquet} formalism~\cite{ts.ok.08,sc.mo.02u,jo.fr.08}
\begin{equation}\label{eq:FloquetGF}
\kel{G}_{mn}(\omega) =\int \dd t_{\text{rel}} \int_{-\tau/2}^{\tau/2} \frac{\dd t_{\text{av}}}{\tau} \ee^{\ii[\left(\omega+m\Omega\right) t -\left( \omega+n\Omega\right)t^{\prime}]} \kel{G}(t,t^{\prime})
\end{equation}
with the \emph{Wigner} GF~\cite{ts.ok.08}
\begin{equation}\label{eq:WignerGF}
\kel{G}_{l}(\omega) =\int \dd t_{\text{rel}} \int_{-\tau/2}^{\tau/2} \frac{\dd t_{\text{av}}}{\tau} \ee^{\ii\omega t_{\text{rel}} + \ii l\Omega t_{\text{av}}} \kel{G}(t,t^{\prime}) \;.
\end{equation}
As usual, $t_{\text{rel}} = t-t^{\prime}$ and $t_{\text{av}} = (t+t^{\prime})/2$ are the relative and average time variables. Eqs.~\eqref{eq:FloquetGF} and (\ref{eq:WignerGF}) are connected via
\begin{equation}\label{eq:Wig2Fl}
\kel{G}_{mn}(\omega)=\kel{G}_{m-n}\left(\omega+\frac{m+n}{2}\Omega \right).
\end{equation}
In this work any Floquet-represented matrix is referred to as either $X_{mn}$ explicitly showing its indices or by means of a boldface letter $\mat{X}$. On the other hand, a single index as subscript $X_{l}$ is reserved for the Wigner representation. Finally, an underline indicates the Keldysh structure 
\begin{equation}\label{eq:Keld-structure}
\kel{\mat{G}} \equiv 
\begin{pmatrix}
\mat{G}^{\text{R}} & \mat{G}^{\text{K}}\\
\mat{0}         & \mat{G}^{\text{A}} \\
\end{pmatrix},
\end{equation}
which contains the \emph{retarded}, \emph{advanced} and \emph{Keldysh} components $\mat{G}^{\text{R,A,K}}$, where $\mat{G}^{\text{A}}=(\mat{G}^{\text{R}})^{\dagger}$. It is worth recalling the definition $\mat{G}^{\text{K}} \equiv \mat{G}^{>} + \mat{G}^{<}$, with $\mat{G}^{\lessgtr}$ denoting the \emph{lesser} and \emph{greater} components~\cite{schw.61,keld.65,ra.sm.86,ha.ja}.

\subsection{Dyson equation}\label{sec:Dyson-eq}
The lattice Floquet GF obeys the Dyson equation 
\begin{equation}\label{eq:FullDysonEq}
\kel{\mat{G}}^{-1}(\omega,\epsilon,\overline{\epsilon}) = \kel{\mat{G}}^{-1}_{0}(\omega,\epsilon,\overline{\epsilon}) - \kel{\mat{\Sigma}}(\omega,\epsilon,\overline{\epsilon}) - \kel{\mat{\Sigma}}_{\text{e-ph}}(\omega,\epsilon,\overline{\epsilon})
\end{equation}
with both electron and e-ph self-energies depending on the crystal momentum via $\epsilon$, $\overline{\epsilon}$. The lattice GF of the non-interacting part of the Hamiltonian~\eqref{eq:MicroHamiltonian} reads
\begin{equation}\label{eq:non-int_InvGF} 
\kel{G}^{-1}_{0,mn}(\omega,\epsilon,\overline{\epsilon}) =  \kel{g}^{-1}_{0,mn}(\omega,\epsilon,\overline{\epsilon}) - \delta_{mn} \ v^{2} \kel{g}_{\text{bath}}(\omega+n\Omega)
\end{equation}
with 
\begin{align}\label{eq:inv_non-int_lat_GF_comps}
\begin{split}
\left[ g^{-1}_{0}(\omega,\epsilon,\overline{\epsilon}) \right]^{\text{R}}_{mn} & = \left( \omega+n\Omega + i0^{+} -\varepsilon_c \right)\delta_{mn} - \varepsilon_{mn}(\epsilon,\overline{\epsilon}), \\
\left[ g^{-1}_{0}(\omega,\epsilon,\overline{\epsilon}) \right]^{\text{K}}_{mn} & \to 0.
\end{split}
\end{align}
The inverse Keldysh component is negligible in comparison with $\Im (g^{\text{K}}_{\text{bath}})$ since it is proportional to an imaginary infinitesimum. The Floquet dispersion relation $\varepsilon_{mn}$ for a dc field in a hypercubic lattice~\cite{ts.ok.08} is given by	
\begin{equation}\label{eq:Floquet_disp}
\varepsilon_{mn}(\epsilon,\overline{\epsilon}) = \frac{1}{2} \left[ \left( \epsilon + i \overline{\epsilon} \right)\delta_{m-n,1} +  \left( \epsilon - i \overline{\epsilon} \right)\delta_{m-n,-1} \right].
\end{equation}
In the wide-band limit considered here, the \emph{retarded} and \emph{Keldysh} components of the electronic bath's GF~\eqref{eq:non-int_InvGF} \cite{ne.ar.15} read
\begin{align}\label{eq:WBL_bathGF}
\begin{split}
v^{2}g^{\text{R}}_{\text{bath}}(\omega) & = - \frac{i}{2} \ \Gamma_{\text{e}}, \\
v^{2}g^{\text{K}}_{\text{bath}}(\omega) & = -i \ \Gamma_{\text{e}} \ \tanh\resub{\left[ \frac{\beta}{2}\left(\omega-\mu\right)\right]},
\end{split}
\end{align}
where the \emph{Keldysh} component is obtained from the \emph{fluctuation-dissipation} theorem $g^{\text{K}}_{\text{bath}}(\omega) = \left[ g^{\text{R}}_{\text{bath}}(\omega) - g^{\text{A}}_{\text{bath}}(\omega) \right] \tanh\left( \beta\left(\omega-\mu\right)/2\right)$. In Eq.~\eqref{eq:WBL_bathGF}, \resub{$\mu$ denotes the chemical potential, while} $\beta \equiv 1/T$ is the inverse temperature and $\Gamma_{\text{e}}$ the coupling strength to the fermionic reservoir. The electron self-energy (SE) $\kel{\mat{\Sigma}}$ is obtained from F-DMFT the only approximation of which consists in neglecting its crystal momentum dependence $\kel{\mat{\Sigma}}(\omega,\epsilon,\overline{\epsilon}) \approx \kel{\mat{\Sigma}}(\omega)$. Further details will be given in Sec.~\ref{sec:FDMFT_implementation}.

\subsubsection{Implementation of the electron-phonon SE}
Within the F-DMFT approximation, the e-ph SE is also included as a local contribution $\kel{\mat{\Sigma}}_{\text{e-ph}}(\omega,\epsilon,\overline{\epsilon}) \approx \kel{\mat{\Sigma}}_{\text{e-ph}}(\omega)$. In terms of the contour time arguments $z,z^\prime$ it has the form
\begin{equation}\label{eq:backbone_e-ph_SE}
\Sigma_{\text{e-ph}}(z,z^{\prime}) = ig^{2} G(z,z^{\prime}) D_{\text{ph}}(z,z^{\prime})
\end{equation}
corresponding to the lowest-order diagram in the bare phonon propagator $D_{\text{ph}}$. The explicit form of the retarded and Keldysh components associated with Eq.~\eqref{eq:backbone_e-ph_SE} can be found in Appendix~\ref{sec:real-time_eph_se}. In Eq.~\eqref{eq:backbone_e-ph_SE}, $G(z,z^{\prime})$ is the local electron Green's function, the Keldysh-Floquet representation of which is obtained as
\begin{equation}\label{eq:Lat_LocGF}
\begin{split}
\kel{\mat G}_{\text{loc}}(\omega) &= \int \dd\epsilon \int \dd\overline{\epsilon} \ \rho(\epsilon,\overline{\epsilon}) \times \\ 
&\times \left[ \left( \kel{\mat G}^{-1}_{0}(\omega,\epsilon,\overline{\epsilon}) - \kel{\mat\Sigma}(\omega) - \kel{\mat \Sigma}_{\text{e-ph}}(\omega) \right)^{-1} \right].
\end{split}
\end{equation}	
Due to gauge invariance, $\kel{\mat G}_{\text{loc}}(\omega)$ is diagonal in Floquet indices in the case of a dc-field~\cite{ts.ok.08}. The Keldysh components~\cite{ao.ts.14} of $D_{\text{ph}}(z,z^{\prime})$ at equilibrium~\footnote{\resub{Feedback-effects from electrons onto phonons are neglected within the {\em non-self-consistent} version of the Migdal approximation employed in this work. Thus phonons cannot heat up.}} are given by
\begin{align}\label{eq:Ph_Prop_time}
\begin{split}
D^{\text{R}}_{\text{ph}}(t,t^{\prime}) & = -i \theta(t-t^{\prime}) \int \dd\omega \ \ee^{-i\omega\left(t-t^{\prime}\right)} A_{\text{ph}}(\omega), \\
D^{>}_{\text{ph}}(t,t^{\prime}) & = -i \int \dd\omega \ \ee^{-i\omega\left(t-t^{\prime}\right)} A_{\text{ph}}(\omega) \left( 1 + b(\omega) \right) \\
D^{<}_{\text{ph}}(t,t^{\prime}) & = -i \int \dd\omega \ \ee^{-i\omega\left(t-t^{\prime}\right)} A_{\text{ph}}(\omega) \ b(\omega), \\
\end{split},
\end{align}
where $b(\omega)=1/(\ee^{\beta\omega}-1)$ is the Bose-Einstein distribution function at inverse temperature $\beta$. We focus on acoustic phonons, with spectral function $A_{\text{ph}}(\omega) = \omega/(4\omega^{2}_{\text{ph}}) \ee^{-| \omega|/\omega_{\text{ph}}}$, $\omega_{\text{ph}}$ being a soft cutoff frequency~\cite{pi.li.21}. Details about the explicit form of the phonon spectral function can be found in Appendix~\ref{sec:phonon_spectrum}.

\subsection{Floquet DMFT}\label{sec:FDMFT_implementation}
The electron SE in Eq.~(\ref{eq:FullDysonEq}) is determined within the DMFT approach~\cite{me.vo.89,ge.ko.92,ge.ko.96}, and in particular its non-equilibrium Floquet (F-DMFT) extension~\cite{ts.ok.08,sc.mo.02u,jo.fr.08}. This consists in neglecting the crystal momentum dependence of the electron SE, i.e. $\kel{\mat{\Sigma}}(\omega,\epsilon,\overline{\epsilon}) \to \kel{\mat{\Sigma}}(\omega)$. This reduces to mapping the original problem onto a single-site impurity model --- the GF $\kel{\mat{g}}^{-1}_{0,\text{site}}(\omega)$ of which is defined as in Eq.~\eqref{eq:inv_non-int_lat_GF_comps} by replacing $\varepsilon_{mn}(\epsilon,\overline{\epsilon}) \to \varepsilon_{\text{c}}$ --- encoding the effect of all other lattice sites into an \emph{effective} bath hybridization function, $\kel{\mat\Delta}(\omega)$. 

The GF of the resulting impurity model must be determined by a many-body impurity solver, in this case a non-equilibrium one, which constitutes the bottleneck of the approach. The hybridization $\kel{\mat\Delta}(\omega)$ is fixed by the condition that the local and impurity GFs coincide, i.e. $\kel{\mat{G}}_{\text{imp}}(\omega)\overset{!}{=}\kel{\mat{G}}_{\text{loc}}(\omega)$. In practice, this self-consistency condition is carried out by (i) starting from a guess for the two SEs $\kel{\mat{\Sigma}}(\omega)$ and $\kel{\mat{\Sigma}}_{\text{e-ph}}(\omega)$. (ii) Computing the GF and obtaining $\kel{\vec{G}}_{\text{loc}}(\omega)$ following Eq.~\eqref{eq:Lat_LocGF}. (iii) Computing $\kel{\mat{\Sigma}}_{\text{e-ph}}(\omega)$ from ~\eqref{eq:backbone_e-ph_SE} (see also Appendix~\ref{sec:real-time_eph_se}). (iv) Extracting  $\kel{\mat\Delta}(\omega)$ from 
\begin{equation}\label{eq:imp_Dyson_eq}
\kel{\mat{G}}^{-1}_{\text{loc}}(\omega) = \kel{\mat{g}}^{-1}_{0,\text{site}}(\omega) - \kel{\mat{\Delta}}(\omega) - \kel{\mat{\Sigma}}(\omega).
\end{equation}
(v) Solving the many body impurity problem and determining the new $\kel{\mat{\Sigma}}(\omega)$. (vi) Using the obtained SEs in step (ii), steps (ii)-(v) are then iterated until convergence.

Notice that for the same reason as for the local Green's function, also $\kel{\mat{\Delta}}(\omega)$, $\kel{\mat{\Sigma}}(\omega)$ and $\kel{\mat{\Sigma}}_{\text{e-ph}}(\omega)$ are diagonal in Floquet indices. This, combined with the translation property $\kel{G}_{mm}(\omega) = \kel{G}_{00}(\omega+m\Omega)$ following Eq.~\eqref{eq:FloquetGF} and being valid for the above quantities as well, allows us to restrict the problem to the $(0,0)$-element only.
	
\subsubsection{Auxiliary master equation approach}\label{eq:amea}
As mentioned above, the most difficult part of the DMFT algorithm is the solution of the impurity problem. At the moment, in contrast to equilibrium there is no well established impurity solver for the non-equilibrium case~\cite{ao.ts.14}. The auxiliary master equation approach (AMEA)~\cite{ar.kn.13,do.nu.14,do.ga.15,do.so.17} adopted here maps the impurity problem, whose bath hybridization function is defined as in Eq.~\eqref{eq:imp_Dyson_eq}, onto an \emph{auxiliary} open quantum system consisting of a finite number of bath sites $N_{\text{B}}$ attached to Markovian reservoirs described by the Lindblad equation. Its corresponding non-interacting hybridization function $\kel{\Delta}_{\text{aux}}(\omega)$ is obtained by fitting the original interacting DMFT one. Convergence is reached once $\kel{\Delta}_{\text{aux}}$ agrees with $\kel{\Delta}$ with sufficient accuracy. With the system parameters obtained from $\kel{\Delta}_{\text{aux}}$, the auxiliary problem can then be solved exactly by many-body-diagonalization methods for open quantum systems yielding the required impurity Green's function. The convergence of AMEA increases exponentially with the number of bath sites $N_{\text{B}}$ \cite{do.so.17}. 
To speed up the whole process, it turns out convenient to first solve the auxiliary impurity problem with $N_{\text{B}}=4$ to get suitable initial SEs to be provided as an input for the simulations with $N_{\text{B}}=6$.~\footnote{Notice that the accuracy achieved for a certain $N_{\text{B}}$ with AMEA roughly corresponds to the accuracy obtained with a conventional exact-diagonalization impurity solver with more than twice the number of bath sites. This comparison can be done only in equilibrium. For a non-equilibrium steady-state a conventional exact diagonalization approach is not possible.}

\subsection{Physical quantities}
\label{sec:observables}

The local electronic spectral function is given by 
\begin{equation}\label{eq:local_spec_func}
A(\omega)=-\frac{1}{\pi} \Im[G^{\text{R}}_{\text{loc}}(\omega)],
\end{equation}
where $G^{\text{R}}_{\text{loc}}$ is the \emph{retarded} component of the GF given in Eq.~\eqref{eq:Lat_LocGF}. 

The non-equilibrium distribution function
\begin{equation}\label{eq:NEFD-dist}
F_{\text{neq}}(\omega) = \frac{1}{2} \left\{1 - \frac{1}{2}\frac{\Im[G^{\text{K}}_{\text{loc}}(\omega)]}{\Im[G^{\text{R}}_{\text{loc}}(\omega)]} \right\}
\end{equation}
is related to the electronic spectral occupation function via
\begin{equation}\label{eq:Filling_func}
N_{\text{e}}(\omega) \equiv A(\omega)F_{\text{neq}}(\omega).
\end{equation}

In our units, the time-resolved current $j(t)$ and kinetic energy $E_{\text{kin}}(t)$ are  given by
\begin{align}
j(t) & = 2i \sum_{\vec{k}} [\vec{e}_{0} \cdot \vec{v}_{\vec{k}}(t)] G^{<}_{\vec{k}}(t,t), \label{eq:time_current} \\
E_{\text{kin}}(t) & = -2i \sum_{\vec{k}} \varepsilon_{\vec{k}}(t) G^{<}_{\vec{k}}(t,t), \label{eq:time_energy}
\end{align}
where the group velocity $\vec{v}_{\vec{k}}(t) = \vec{\nabla}_{\vec{k}}\varepsilon_{\vec{k}}(t)$ is obtained from the dispersion relation $\varepsilon_{\vec{k}}(t)=\epsilon \cos[A(t)]+\overline{\epsilon} \sin[A(t)]$ following from Eqs.~\eqref{eq:d-dim_crystal_dep} and \eqref{eq:Floquet_disp}, while the factor $2$ accounts for spin-degeneracy. The Wigner-represented current is given by 
\begin{equation}\label{eq:general_Wig_current}
j_{l}=\int_{-\infty}^{+\infty} \frac{\dd\omega}{2\pi} \ j_{l}(\omega)
\end{equation}
with the frequency-resolved current
\begin{equation}\label{eq:Current_integrald_L}
j_{l}(\omega) = \int \dd\epsilon \int \dd\overline{\epsilon} \ \rho(\epsilon,\overline{\epsilon}) \left[ \left( \epsilon - i\overline{\epsilon} \right) G^{<}_{l+1}(\omega,\epsilon,\overline{\epsilon}) + \text{h.c.} \right].
\end{equation}

The Wigner-represented kinetic energy is
\begin{equation}\label{eq:general_Wig_energy}
E_{\text{kin},l}=\int_{-\infty}^{+\infty} \frac{\dd\omega}{2\pi} \ E_{\text{kin},l}(\omega)
\end{equation} 
with the frequency-resolved kinetic energy
\begin{equation}\label{eq:Energy_integrald_L}
\begin{split}
E_{\text{kin},l}(\omega) &= \int \dd\epsilon \int \dd\overline{\epsilon} \ \rho(\epsilon,\overline{\epsilon}) \times \\
&\times \left[ - \left( \overline{\epsilon} + i\epsilon \right) G^{<}_{l+1}(\omega,\epsilon,\overline{\epsilon}) + \text{h.c.} \right].
\end{split}
\end{equation}	 
The explicit derivation of the expressions above is contained in Appendix~\ref{sec:ss_j_Ekin_derivation}. Again, the $l\neq 0$ components of these quantities vanish by means of gauge invariance due to the time-independent nature of the dc field setup. For this reason, we will drop the index $l$ from now on.

\section{Results}\label{sec:results}
Our aim is to study the effect of a strong dc electric field which is comparable to the gap in Mott insulators. Therefore we choose $U/t^{\ast}=8$ for which the system shows a well established insulating phase~\cite{mu.we.18}. All simulations are carried out \resub{at half-filling with $\varepsilon_{\text{c}}=-U/2$}, inverse temperature $\beta=20/t^{\ast}$, \resub{and chemical potential $\mu=0$}. We define the phonon coupling strength~\footnote{Note that $\Gamma_{\text{ph}}$ cannot be used to infer the phonon relaxation time as $\tau^{-1}_{\text{ph}}\sim \Gamma_{\text{ph}}$. In fact, such an information is encoded in $\Im(\Sigma^{\text{R}}_{\text{e-ph}})$ rather than in $A_{\text{ph}}$. For this reason one should choose $\Gamma_{\text{ph}} = -2\Im \left[\Sigma^{\text{R}}_{\text{e-ph}}(\omega^{\ast})\right]$, with $\omega^{\ast}$ as some significant frequency. However, given the strong dependence of the e-ph SE on both the applied field and frequency, see Figs.~\ref{fig:EQ_PH_SF_PH_SE} and \ref{fig:PH_SFs_maxima}(c)-(d), there is no unique choice of $\omega^{\ast}$, which is why we resort to Eq.~(\ref{eq:coupling_strengths}).} as  
\begin{equation}\label{eq:coupling_strengths}
\Gamma_{\text{ph}} \equiv 2\pi g^{2} A_{\text{ph}}(\omega_{\text{ph}}) = \frac{\pi}{2e} \left( \frac{g^{2}}{\omega_{\text{ph}}} \right),
\end{equation} 
which allows us to introduce the dimensionless electron-to-phonon ratio $\gamma \equiv \Gamma_{\text{e}}/\Gamma_{\text{ph}}$ as a measure of their relative strength, with $\Gamma_{\text{e}}$ given in Eq.~\eqref{eq:WBL_bathGF}. In this paper, we focus on $\gamma \ll 1$ to keep the influence of the fermionic bath as small as possible~\footnote{\resub{We point out that in the sense specified in~\cite{Note1}, the smallest coupling $\Gamma_{\text{e}}$ that ensures convergence over the whole range $F/t^{\ast}\in [0,10]$ equals $\Gamma_{\text{e}}/t^{\ast}=0.065$, for the parameter setup addressed in this work.}}. We choose as default values $\Gamma_{\text{ph}}/t^{\ast}=0.925$ and $\omega_{\text{ph}}/t^{\ast}=0.1$ if not stated otherwise.

As discussed above, phonons provide a valid dissipation channel with a stable steady-state only around specific values of the electric field $\vec{F}$. To achieve numerical stability for a wide range of $\vec{F}$, we need to include a weakly coupled fermionic reservoir. In Sec.~\ref{sec:WBL_plus_ph}, we then vary the coupling $\Gamma_{\text{e}}$ with the fermionic bath in addition to dissipation via phonons. We discuss the case of purely phononic dissipation $\Gamma_{\text{e}}=0$~\footnote{\resub{Phonon-only dissipation does not always provide a stable steady-state in the sense of~\cite{Note1}. To address the case $\Gamma_{\text{e}}=0$ at the two main resonances, we initialize our DMFT runs with the converged SEs of the smallest $\Gamma_{\text{e}}$ employed in Sec.~\ref{sec:WBL_plus_ph}, i.e. $\Gamma_{\text{e}}/t^{\ast}=0.065$ (see also~\cite{Note7}). The electric field strength for this setup is then varied with increments of $\delta F/t^{\ast} =0.02$ around $F \simeq U/2$ and $F\simeq U$.}} in Sec.~\ref{sec:ph_only}.

\subsection{Phonons and electron baths}\label{sec:WBL_plus_ph}

\subsubsection{Spectral properties}\label{sec:e+ph+spectra}
\begin{figure}[t]
\includegraphics[width=0.9\linewidth]{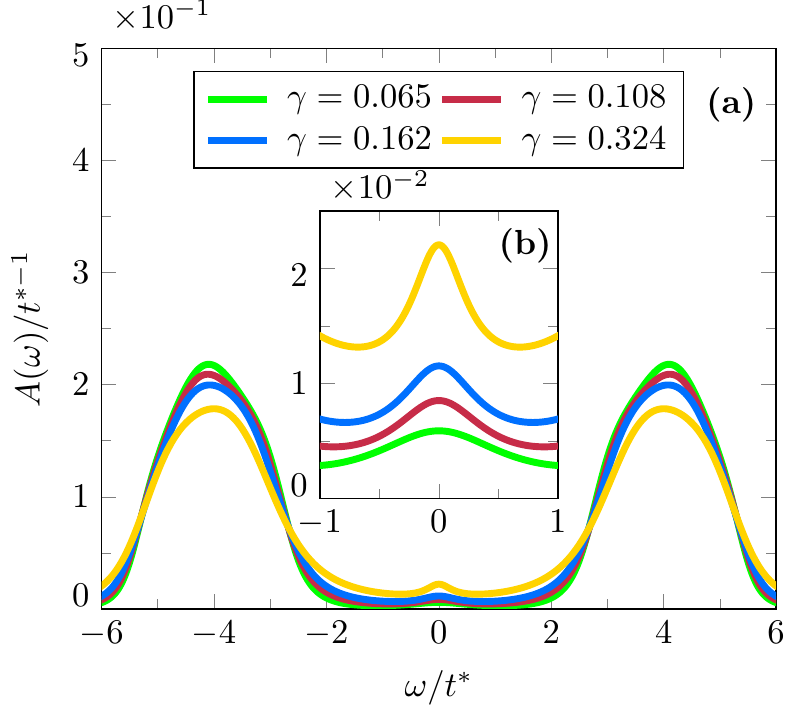}
\caption{(Color online) (a) Spectral function $A(\omega)/t^{\ast-1}$ at zero electric field $F=0$ for different values of the ratio $\gamma$. The inset (b) shows the magnified quasi-particle peak at $\omega=0$. Default parameters are specified at the beginning of section~\ref{sec:results}.}
\label{fig:SFs_QPPePH}
\end{figure}

In equilibrium at $F=0$, the local spectral function $A(\omega)$ shown in Fig.~\ref{fig:SFs_QPPePH}(a) displays the spectrum of the Mott insulator weakly connected to a metallic bath which is characterized by upper and lower Hubbard bands (UHB, LHB) at $\omega_{\text{HB},\pm} \approx \pm U/2$ with bandwidths~\footnote{The bandwidths refer to the full width at half maximum.} of about $2t^{\ast}$.

\begin{figure}[b]
\includegraphics[width=1.0\linewidth]{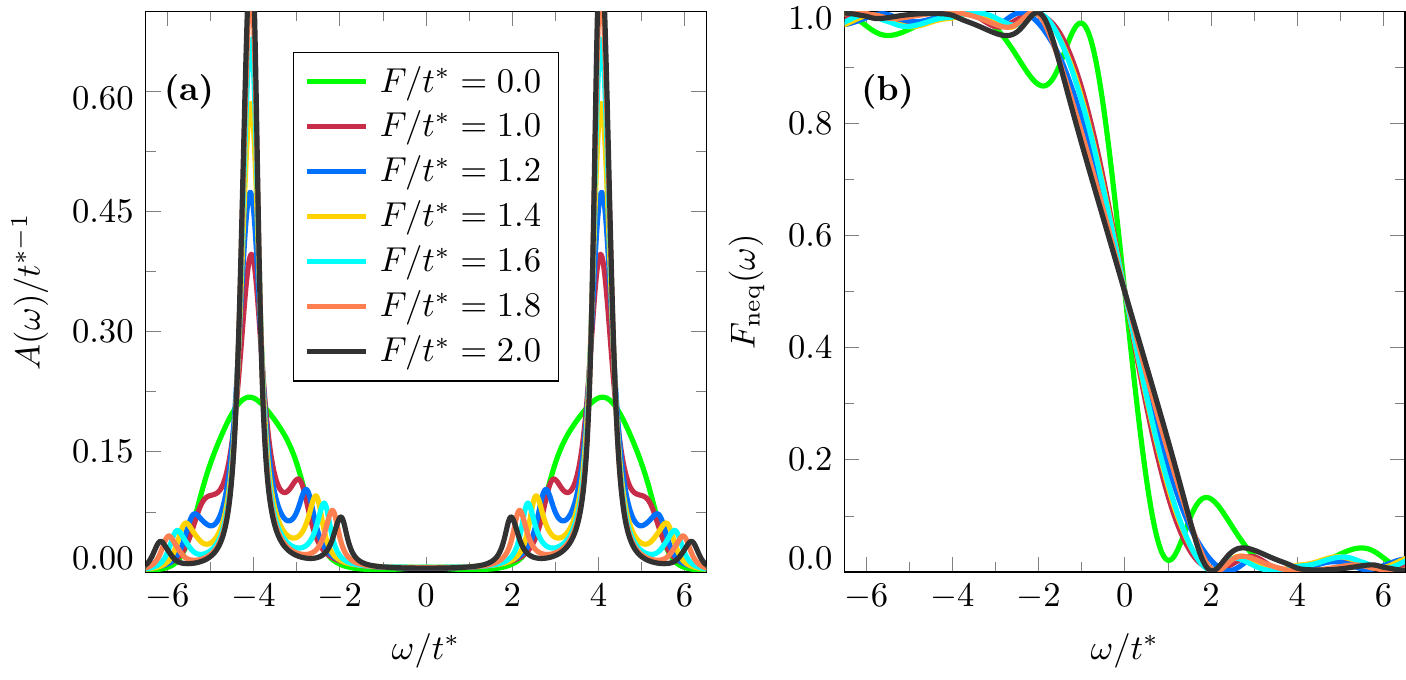}
\caption{\resub{(Color online) (a) Spectral function $A(\omega)/t^{\ast-1}$ and (b) non-equilibrium distribution function $F_{\text{neq}}(\omega)$ for various field strengths at $\gamma=0.065$. Default parameters are specified at the beginning of section~\ref{sec:results}.}}
\label{fig:small_fields_SF_NeqFD}
\end{figure}

\resub{In the regime of small electric fields, increasing the field raises the potential energy drop between neighboring sites, thus localizing the electrons further. The effect is clearly visible in Fig.~\ref{fig:small_fields_SF_NeqFD}(a), where the spectral weight is shifted from the main Hubbard bands to satellite peaks located at integer multiples~\cite{aron.12} of the field $F$. The overall structure is known as Wannier-Stark (WS) ladder in non-interacting systems~\cite{da.wi.88,tu.fr.05} while in the context of many-body physics~\cite{mu.we.18} it has been also referred to as Bloch-Zener archipelago~\cite{aron.12}. Injecting energy into the system via the applied field heats up the electrons, which results in a non-equilibrium distribution function characterized by a higher temperature as compared to the equilibrium case $F=0$, see Fig.~\ref{fig:small_fields_SF_NeqFD}(b).}

\begin{figure}[t]
\includegraphics[width=\linewidth]{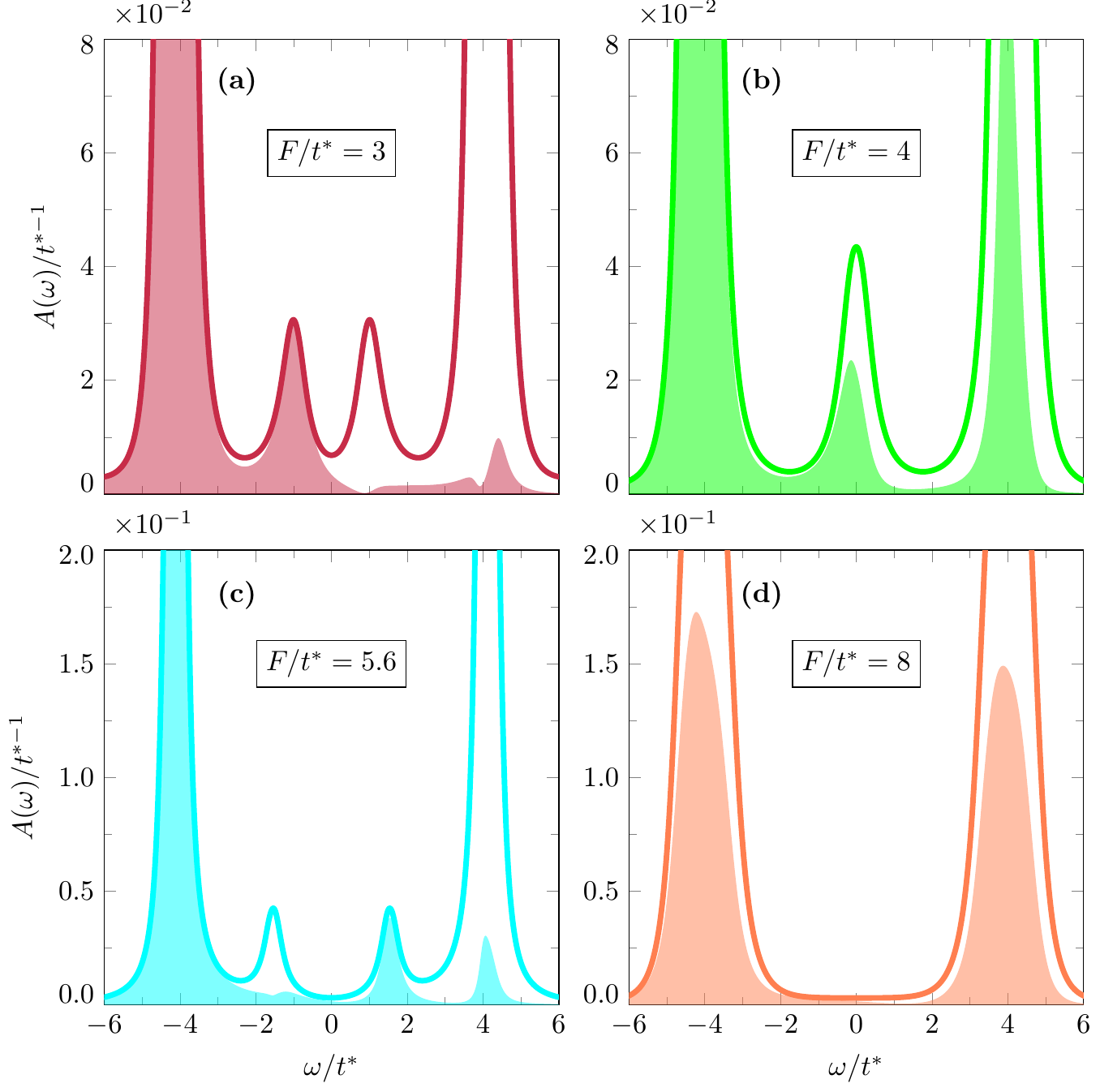}
\caption{(Color online) Spectral function $A(\omega)/t^{\ast-1}$ (solid) and spectral occupation \resub{$N_{\text{e}}(\omega)/t^{\ast-1}$} (shaded area) at (a) $F=3t^{\ast}$, (b) $F=4t^{\ast}$, (c) $F=5.6t^{\ast}$ and (d) $F=8t^{\ast}$. Here $\gamma=0.065$. Default parameters are specified at the beginning of section~\ref{sec:results}.}
\label{fig:SF+Fill_gamma0065}
\end{figure}

At larger fields as shown in Fig.~\ref{fig:SF+Fill_gamma0065} the in-gap WS satellite peaks $\omega_{\text{WS},\pm,\mp}=\pm (U/2) \mp F$ grow in magnitude (cf. Fig.~\ref{fig:small_fields_SF_NeqFD}(a)). In this regime, increasing the field lets these peaks come closer until they overlap at $F\approx U/2$ as evidenced by Fig.~\ref{fig:SF+Fill_gamma0065}(b). Exceeding $U/2$, the electric field pushes the WS peaks further towards the Hubbard bands at $\omega_{\text{HB},\mp}$ as can be inferred from Fig.~\ref{fig:SF+Fill_gamma0065}(c). 

Altogether, the emergence of WS side-bands can be understood as the overlap between LHB and UHB of neighboring sites along the $\vec{e}_{0}$-direction of the electric field. Due to the field-induced potential drop, the Hubbard bands shift in opposite directions until field strengths of $F\approx U$ are reached where the LHB and UHB of neighboring sites overlap, see Fig.~\ref{fig:SF+Fill_gamma0065}(d). The in-gap WS \emph{islands} effectively lower the energy separation between the main Hubbard bands and facilitate the motion of electrons for $F<U$, as we will discuss in the following.

In Fig.~\ref{fig:SF+Fill_gamma0065}(a), we observe that for $F=3t^{\ast}$ the WS side-bands lie at $\omega_{\text{WS},\pm,\mp} \approx \pm t^{\ast}$. Subject to the field, an electron within the LHB gains the energy to reach $\omega_{\text{WS},-,+}\approx -t^{\ast}$. The corresponding spectral occupation $N_{\text{e}}(\omega)$ shows that this side-band is completely occupied. Due to the charge motion from the LHB to $\omega_{\text{WS},-,+} \approx -t^{\ast}$, we expect the current to be slightly enhanced with respect to $F=0$. 
At $F/t^{\ast}=4$, the in-gap WS bands are located at $\omega_{\text{WS},\pm,\mp} \approx 0$, thus making states available around $\omega=0$ that could in principle be filled by the fermionic bath with a negligible cost in energy~\footnote{Due to the presence of the electric field, there is no true overall Fermi energy. This should be seen as the local one.}. However, these states provide the pathway to promote only the electrons belonging to the LHB into the UHB, as will be confirmed by the discussion in Sec.~\ref{sec:ph_only}. In particular, $N_{\text{e}}(\omega)$ shows that the overlapping WS peaks $\omega_{\text{WS},\pm,\mp}$ and the UHB are now considerably occupied. However, the occupation of the former is due to the LHB of one site entering the gap of a neighboring site, while the latter originates from the meeting of the LHB and UHB of sites lying two lattice spacings apart~\cite{mu.we.18,li.ar.15}.  The situation for $F\gtrsim U/2$ can be understood by considering the case $F=5.6t^{\ast}\approx 2U/3 $ in Fig.~\ref{fig:SF+Fill_gamma0065}(c): here the WS side-bands occur at $\omega_{\text{WS},\pm,\mp}\approx \mp1.4t^{\ast}$ and the spectral occupation $N_{\text{e}}(\omega)$ shows that only $\omega_{\text{WS},\pm,\mp}\approx1.4t^{\ast}$ is considerably occupied. As we will see in Sec.~\ref{sec:observables_bothbaths}, also here the current is enhanced due to charge migration from the LHB to the WS side-bands  but it is still smaller than  at the \emph{resonance} $F\approx U$. For $F\approx U$  the spectral function shows no WS peaks because of the perfect matching of the lower and upper Hubbard bands. At the same time $N_{\text{e}}(\omega)$ shows that they are almost equally filled, Fig.~\ref{fig:SF+Fill_gamma0065}(d). We note that this condition is the signature of the fermionic bath draining particles from the UHB back to the LHB \cite{aron.12}. A comparison with the case including only phonons as the source of dissipation discussed in Sec.~\ref{sec:ph_only} will corroborate this statement. 

Second order WS side-bands at $\omega_{\text{WS}2,\pm,\mp 2}=\pm (3U/2)\mp 2F$ can be also observed, confirming the findings in Ref.~\cite{mu.we.18}.
\begin{figure}[t]
\includegraphics[width=\linewidth]{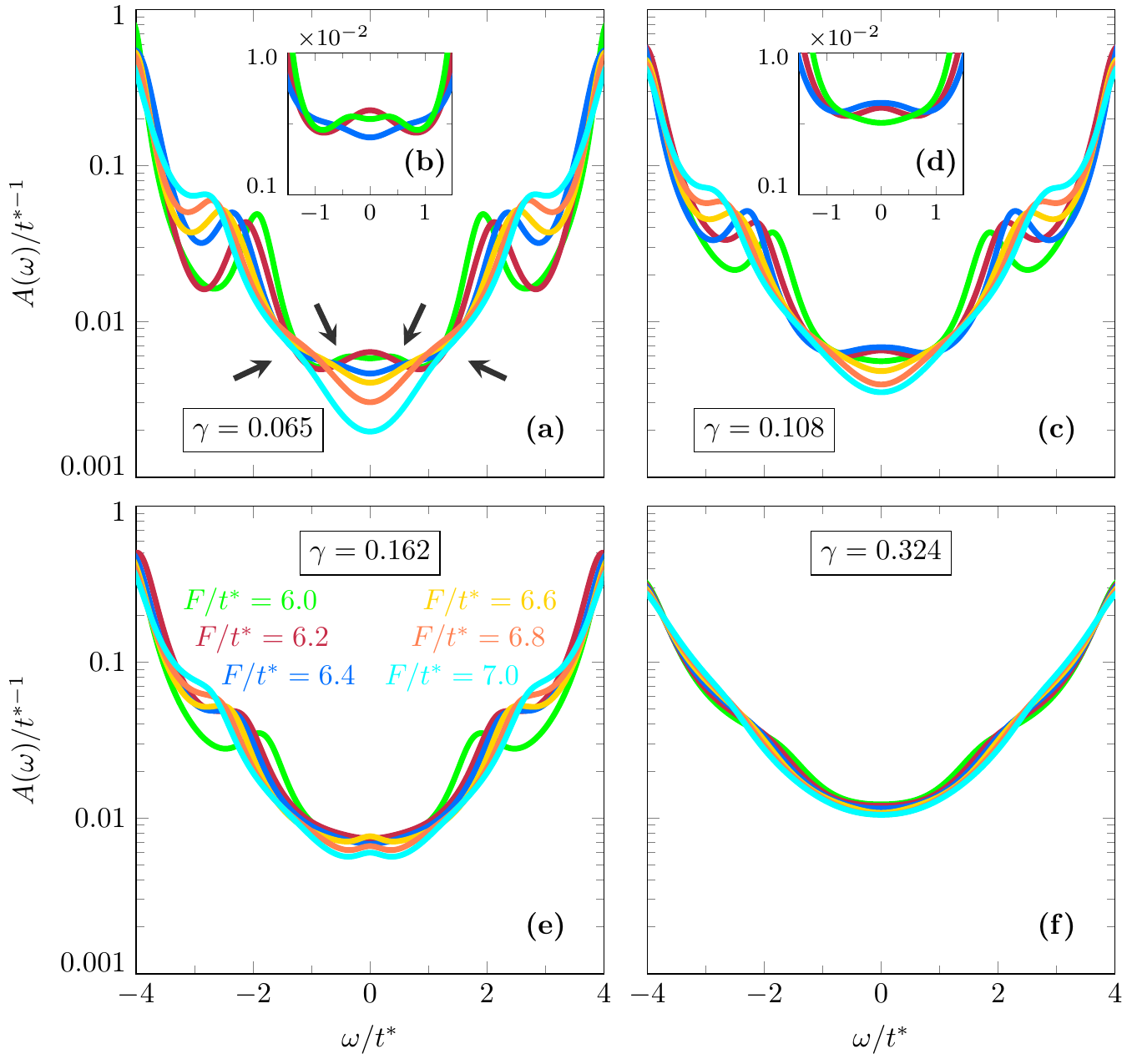}
\caption{(Color online) \resub{(a) WS in-gap states at $\omega_{\text{WS},\pm,\mp}=\pm (U/2) \mp F$ in the spectral function $A(\omega)/t^{\ast-1}$, for various fields $F$ at $\gamma=0.065$. Black arrows point to the position of $\omega_{\text{WS2},\pm,\mp2}=\pm (3U/2) \mp 2F$, while inset (b) shows the magnified region $\omega/t^{\ast} \in [-1.5,1.5]$ for selected field strengths. [(c),(d)] Same for $\gamma=0.108$. [(e),(f)] WS in-gap states $\omega_{\text{WS},\pm,\mp}$ for $\gamma=0.162$ and $\gamma=0.324$.} Default parameters are specified at the beginning of section~\ref{sec:results}.}
\label{fig:SFs_EPH_strengths}
\end{figure}
\resub{These faint features, highlighted by black arrows in Fig.~\ref{fig:SFs_EPH_strengths}(a), first move towards one another as the field increases (see the inset Fig.~\ref{fig:SFs_EPH_strengths}(b) for $F/t^{\ast}=\left\{ 6.0, 6.2, 6.4 \right\}$) and then run past each other, approaching the main Hubbard bands as the field is further increased.} They are associated with adding a particle to a single occupied site (energy $U/2$) with the simultaneous creation of a doublon-holon pair (energy $U$), accompanied by a tunneling of two particles over one lattice site. \resub{However accounting for second order processes, these side-bands become relevant at $F\gtrsim U/2$ but are only visible on the logarithmic scale as shown \resub{in Figs.~\ref{fig:SFs_EPH_strengths}(a)-(b)}. Also, they are easily washed away by increasing the coupling $\Gamma_{\text{e}}$ to the fermionic bath, as evidenced in Figs.~\ref{fig:SFs_EPH_strengths}(c)-(f).} To visualize their faint signature, consider the case of $F=5.6t^{\ast}\approx2U/3$ shown in Fig.~\ref{fig:SF+Fill_gamma0065}(c). Here the WS states $\omega_{\text{WS}2,\pm,\mp 2}\approx \pm0.8t^{\ast}$ do not contribute significantly to the current: as evidenced from $N_{\text{e}}(\omega)$, the net motion of charge carriers mainly involves the transition $\omega_{\text{HB},-}\approx -U/2 \rightarrow \omega_{\text{WS},-,+}\approx -U/2 + 2U/3$, in agreement with previous work~\cite{mu.we.18}.

The spectral properties described so far are quite robust against the inclusion of phonons as can be inferred by comparison with Ref.~\cite{mu.we.18} which only incorporated dissipation via electrons. We note however, that increasing $\Gamma_{\text{e}}$ tends to fill the gap, as can be seen both in the inset Fig.~\ref{fig:SFs_QPPePH}(b) as well as in Figs.~\ref{fig:SFs_EPH_strengths}(c)-(f). This is accompanied by an evident trend to \emph{wash away} the WS satellite peaks. For a given field strength $F$, a larger $\Gamma_{\text{e}}$ flattens the states $\omega_{\text{WS}2,\pm,\mp 2}$ and broadens the sub-peaks $\omega_{\text{WS},\pm,\mp}$, accelerating their merging with the Hubbard bands. WS states are then barely visible even on a logarithmic scale as is the case for $\gamma=0.324$ seen in Fig.~\ref{fig:SFs_EPH_strengths}(f).

\resub{Finally, a brief remark on the orientation of the electric field and how it affects the results presented thus far. We observe that resonances like the one at $F\simeq U/2$, and their meltdown as the field moves off that value, appear to be quite robust, see also the so-called {\em Bloch-Zener islands} mentioned in Ref.~\cite{aron.12} where $\vec{F} = F \vec{e}_{\text{x}}$. On the other hand, the fact that neighboring sites do not have iso-energy surfaces may prevent or strongly suppress the dimensional crossover~\cite{ar.ko.12,aron.12} at $F\simeq U$ due to the equivalence of all spatial directions.}

\subsubsection{Current, energy and double occupancy}\label{sec:observables_bothbaths}
\begin{figure}[t]
\includegraphics[width=0.8\linewidth]{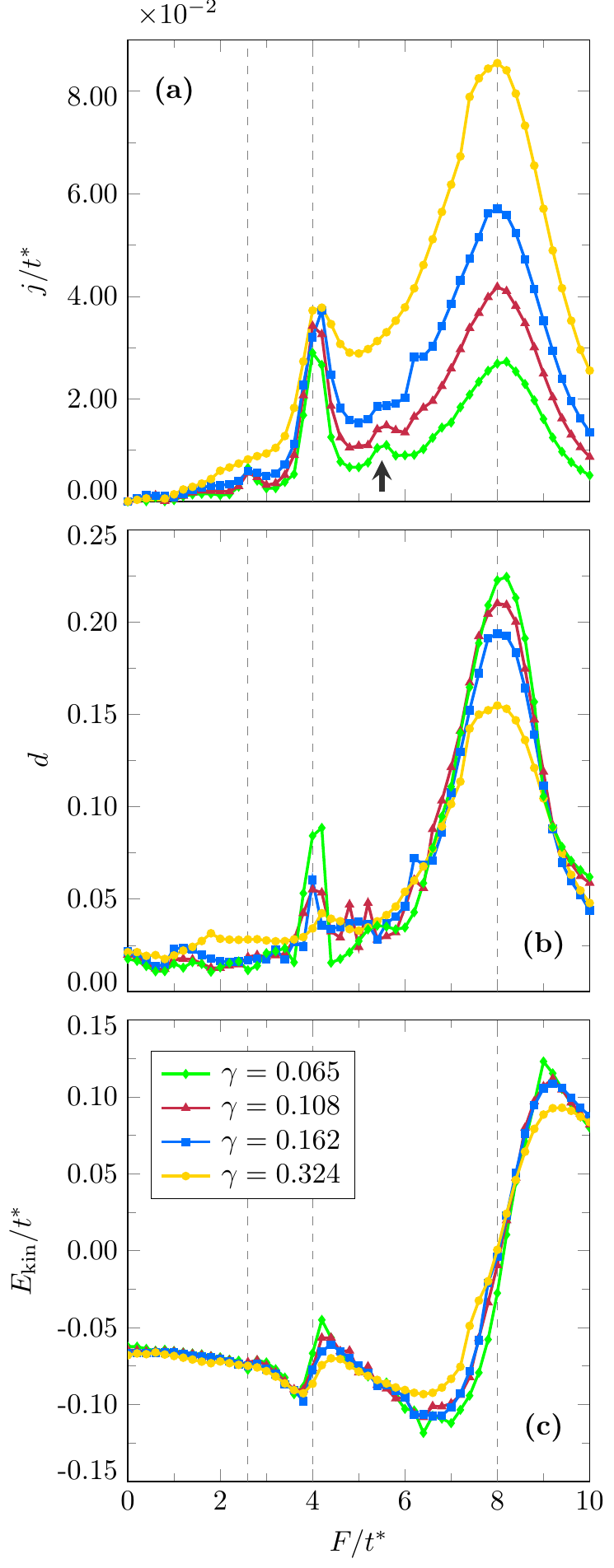}
\caption{(Color online) [(a),(b),(c)] Electric field dependence of the steady-state current $j$, double occupancy $d$, and kinetic energy $E_{\text{kin}}$ for various $\gamma = \Gamma_{\text{e}}/\Gamma_{\text{ph}}$. Dashed vertical lines denote the resonant fields $F=U/3$, $U/2$ and $U$, while the black arrow highlights the resonance at $F\approx5.6t^{\ast}$. Default parameters are specified at the beginning of section~\ref{sec:results}.}
\label{fig:Observables_bothbaths}
\end{figure}

In the Coulomb gauge, an electron that tunnels through $m$ lattice spacings acquires or loses the potential energy $E_{m}(F)=mF$. In particular, when the difference is $\Delta E \equiv U-E_m(F) \approx 0$, the current is enhanced due to the resonant creation of doublon-holon pairs which lie $m$ sites apart. However these resonances are exponentially suppressed by a factor $\left( t^{\ast}/U \right)^{|m|}$.
For instance, Fig.~\ref{fig:Observables_bothbaths}(a) shows clear resonances in the steady-state current~\eqref{eq:general_Wig_current} at $F\approx U/2$ and $F\approx U$, as expected from Sec.~\ref{sec:e+ph+spectra} (cf. Fig. ~\ref{fig:SF+Fill_gamma0065}(b)-(d)).  In these cases charge migration is due to electrons tunnelling from LHB to UHB belonging to sites $|m|=2$ ($F=U/2$) and $|m|=1$ ($F=U$) lattice spacings apart. The peak at $F\approx U/3$, on the other hand, is strongly suppressed according to $\left( t^{\ast}/U \right)^{3}$  as it accounts for tunnelling over $|m|=3$ lattice sites. In any case, in order to sustain a current, resonant tunneling from a LHB to an UHB $m$ sites away must be followed by relaxation to the LHB. For this reason, the current increases upon increasing $\gamma$, as visible in Fig.~\ref{fig:Observables_bothbaths}(a). However, the resonances occurring at $F>U/2$ are significantly more sensitive to a larger $\gamma$ than those at $F<U/2$, the latter being almost unaltered while the former get greatly boosted.

\begin{figure}[t]
\includegraphics[width=\linewidth]{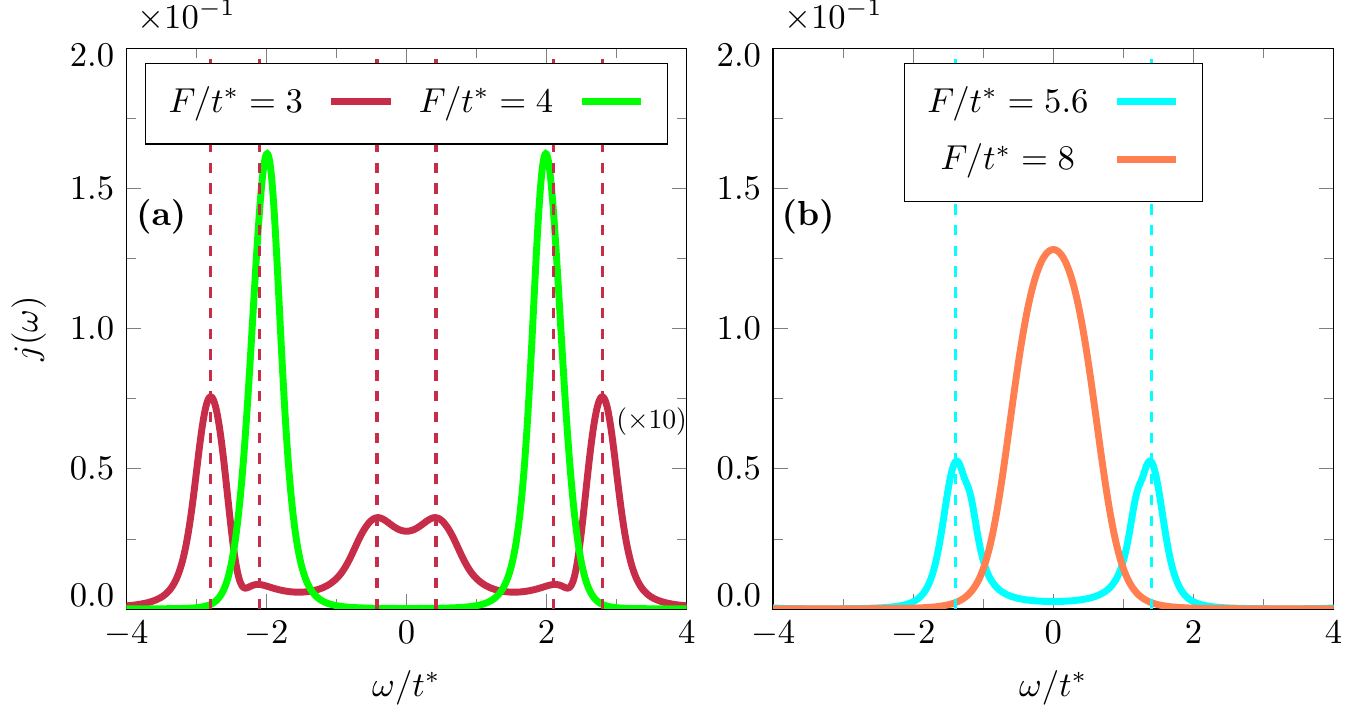}
\caption{(Color online) (a) Frequency resolved current $j(\omega)$~\eqref{eq:Current_integrald_L} at $F=3t^{\ast}$ and $F=4t^{\ast}$. (b) Same quantity shown for $F=5.6t^{\ast}$ and $F=8t^{\ast}$. Dashed vertical lines mark the maxima discussed in the text. Default parameters are specified at the beginning of section~\ref{sec:results}.}
\label{fig:Jw_true_res}
\end{figure}

Fig.~\ref{fig:Jw_true_res} displays the frequency resolved current~\eqref{eq:Current_integrald_L} at field strengths and $\gamma$ values matching those in Fig.~\ref{fig:SF+Fill_gamma0065} for $A(\omega)$. The largest peaks in $j(\omega)$ marked by dashed lines lie at $\omega_{\text{j},\pm,\mp} = \pm U/2 \mp F/2$: for $F=3t^{\ast}$ they are located at $\omega \approx \pm 2.8t^{\ast}$, see Fig.~\ref{fig:Jw_true_res}(a). However, the sub-peaks at $\omega/t^{\ast}\approx \{ \pm 0.4, \pm 2.2 \}$ are not captured by $\omega_{\text{j},\pm,\mp}$. 
Their origin can be understood with the help of the linear response formula for the tunnelling current $j_{\text{tun}}(\omega)$ presented in Ref.~\cite{mu.we.18}, see Eq.~(20) therein. In terms of the spectral occupations, it can be written as  
\begin{equation}\label{eq:jtun}
j_{\text{tun}}(\omega)=\pi t^{\ast 2}\left[ N_{\text{e}}(\omega)N_{\text{h}}(\omega+F) - N_{\text{e}}(\omega+F)N_{\text{h}}(\omega) \right],
\end{equation}
where $N_{\text{h}}(\omega)\equiv A(\omega)\left( 1-F_{\text{neq}}(\omega)\right)$ is the hole spectral occupation function. 

\begin{figure}[t]
\includegraphics[width=\linewidth]{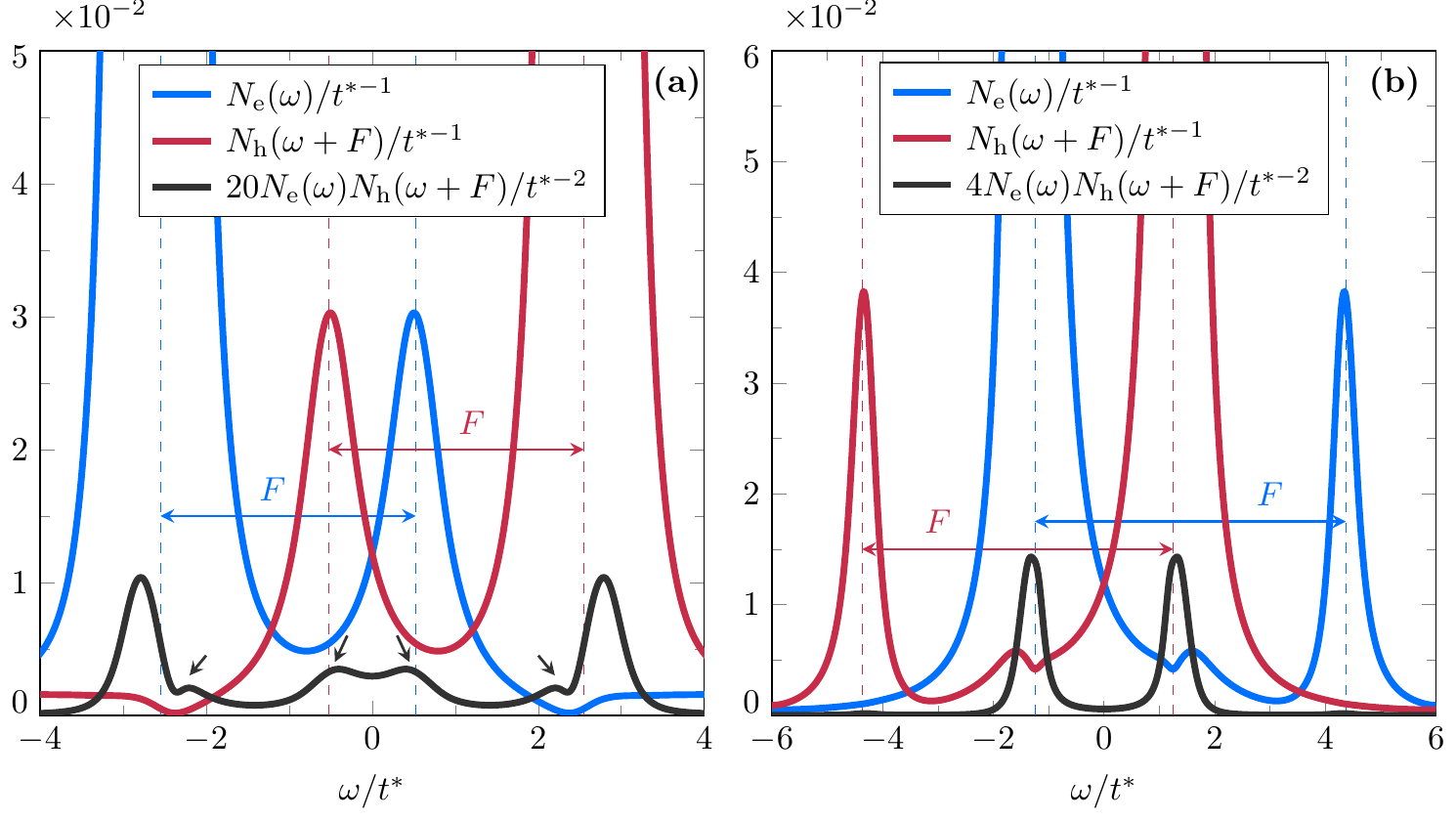}
\caption{(Color online) (a) Electron and hole spectral occupation functions \resub{$N_{\text{e}}(\omega)/t^{\ast-1}$} and \resub{$N_{\text{h}}(\omega+F)/t^{\ast-1}$} and their product at $F=3t^{\ast}$. Black arrows point at the sub-peaks discussed in the text, see also Fig.~\ref{fig:Jw_true_res}(a). (b) Same quantities for $F=5.6t^{\ast}$. \resub{Here $\gamma=0.065$} while default parameters are specified at the beginning of section~\ref{sec:results}.}  
\label{fig:SFs_tunnel}
\end{figure}

Positive contributions to $j_{\text{tun}}(\omega)$ originate from particles flowing from the filled low energy states $N_{\text{e}}(\omega)$ to the empty high energy ones $N_{\text{h}}(\omega+F)$. It is sufficient to analyze the term $N_{\text{e}}(\omega)N_{\text{h}}(\omega+F)$.  In particular, we discuss the cases of $F=3t^{\ast}$ and $F=5.6t^{*}\approx2U/3$. For $F=3t^{\ast}$ shown in Fig.~\ref{fig:SFs_tunnel}(a), the maxima of $N_{\text{e}}(\omega)$ and $N_{\text{h}}(\omega+F)$ do not overlap, resulting in a very small contribution to $j_{\text{tun}}(\omega)$ made clear by the scaling factor of $N_{\text{e}}(\omega)N_{\text{h}}(\omega+F)$. The sub-peaks at $\omega/t^{\ast}\approx \{ \pm 0.4, \pm 2.2 \}$ follow as the result of the overlap of $N_{\text{e}}(\omega)$ and $N_{\text{h}}(\omega+F)$. For $F=5.6t^{\ast}$ displayed in Fig.~\ref{fig:SFs_tunnel}(b), the regions having the largest electron and hole occupations are closer and lead to a hump in the current, see Fig.~\ref{fig:Observables_bothbaths}(a). From these considerations we see that at $F=8t^{*}$ in Fig.~\ref{fig:Jw_true_res}(b), the electrons can now directly fill the vacancies, as evidenced by the single-peak structure of the frequency resolved current $j(\omega)$. At the onset of the resonance this effect is balanced by the term $N_{\text{e}}(\omega+F)N_{\text{h}}(\omega)$ since the UHB contains a considerable fraction of particles in the high energy regions of the spectrum, see Fig.~\ref{fig:SF+Fill_gamma0065}(d).

We emphasize that for $\gamma=0.065$ the steady-state current at $F\approx U/2$ is slightly larger than the one obtained at $F=U$, Fig.~\ref{fig:Observables_bothbaths}(a). This imbalance contrasts with the results in Ref.~\cite{mu.we.18} for the same coupling $\Gamma_{\text{e}}$ but agrees qualitatively with those in Ref.~\cite{le.pa.14}, where a master equation approach employing phononic reservoirs is used as impurity solver. Further details will be given in Sec.~\ref{sec:ph_only}.

The double occupation per site is shown in Fig.~\ref{fig:Observables_bothbaths}(b). Its main resonances at $F\approx U/2$ and $F\approx U$ corroborate the fact that the current is directly related to the creation of doublon-holon pairs. The importance of these processes is evidenced by the resonance at $F\approx 2U/3$ discussed in Sec.~\ref{sec:WBL_plus_ph}, which is clearly visible at $\gamma=0.065$ and gets smeared out with increasing $\gamma$. 

In contrast to the trend exhibited by the current $j$, a stronger fermionic bath suppresses the double occupancy at both $F\approx U/2$ and $F\approx U$. This difference can be explained as follows: as already pointed out~\cite{aron.12}, a larger $\gamma$ increases the number of particles that are drained from the UHB and injected back into the LHB. This, results in a higher annihilation rate of \emph{doublons}, effectively decreasing the fraction of doubly occupied sites within the lattice.

Since it is related to the creation of a doubly occupied site, knowing $d$ provides information on the interaction energy between the electrons. To fully understand the energy balance, knowledge of the kinetic energy $E_{\text{kin}}$~\footnote{We recall that in a simple tight-binding model the kinetic energy starts at negative values.} is required as well. Fig.~\ref{fig:Observables_bothbaths}(c) shows $E_{\text{kin}}$ as a function of the applied field. At low fields, $E_{\text{kin}}$ decreases as the field is increased independently of $\gamma$ and in contrast to both $j$ and $d$. It reaches a local minimum at $F\approx 3.8t^{\ast}$, but turns towards a local maximum past $U/2$ at $F/t^{\ast}\approx 4.2\text{--}4.4$, where $d$ is already showing a downward trend. This behavior signals the energy redistribution within the system. In fact, the field can either enhance the motion of particles, for which an intuitive measure is given by $E_{\text{kin}}$, or the energy can be transferred to particles in a way that allows for a larger number of doubly occupied sites. This interpretation is corroborated by the global maximum of $E_{\text{kin}}$ at $F\approx 9t^{\ast}$, a field strength where $d$ is suppressed.

\subsection{Dissipation by phonons only}\label{sec:ph_only}
\begin{figure}[t]
\includegraphics[width=.9\linewidth,height=0.5\textheight,keepaspectratio]{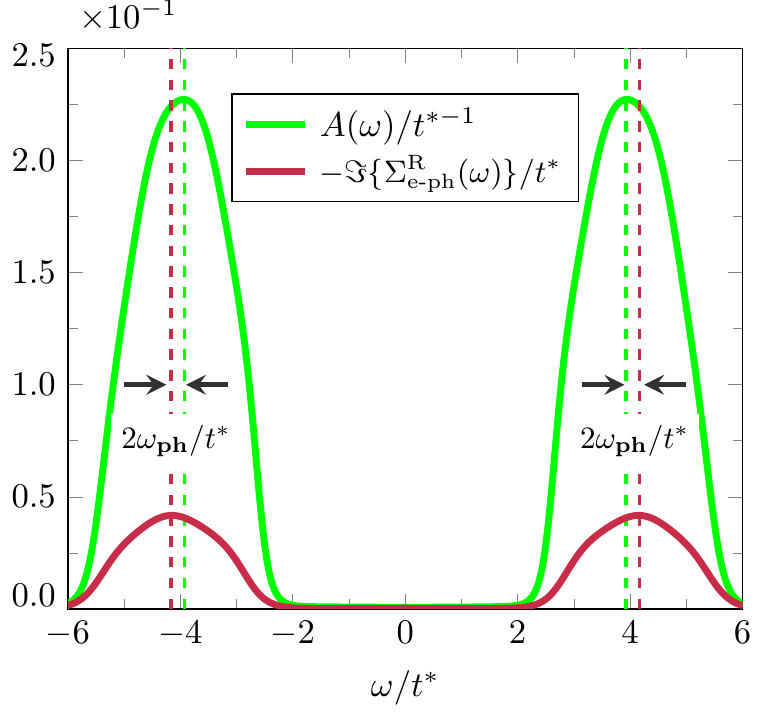}
\caption{(Color online) Electron spectral function $A(\omega)/t^{\ast-1}$ and \resub{e-ph SE} $-\Im[\Sigma^{\text{R}}_{\text{e-ph}}(\omega)]/t^{\ast}$ at $F=0$ for $\gamma=0$. Dashed vertical lines mark the peaks of $A$ and $-\Im(\Sigma^{\text{R}}_{\text{e-ph}})$, while arrows highlight the shifts between \resub{them}. Default parameters are specified at the beginning of section~\ref{sec:results}.}
\label{fig:EQ_PH_SF_PH_SE}
\end{figure}

We now discuss the setup including only phonons as dissipation channel, i.e. $\Gamma_{\text{e}}=0$~\footnote{\resub{Upon removal of the fermionic bath one no longer can explicitly fix the chemical potential of the system. In our setup it is implicitly set by enforcing particle-hole symmetry and using the procedure in~\cite{Note8}.}}. In this case, the e-ph SE $\kel{\Sigma}_{\text{e-ph}}$ is strongly determined by $G_{\text{loc}}(\omega)$. In particular a local GF with a gap generates a gapped e-ph SE with the imaginary part of its \emph{retarded} component vanishing in the same frequency region where $\Im(G^{\text{R}}_{\text{loc}})$ vanishes (cf. Fig.~\ref{fig:EQ_PH_SF_PH_SE}). Thus phonons contribute to the dissipation only within a limited frequency window.

\subsubsection{Spectral properties}\label{sec:PH_SF_Filling}
In equilibrium at zero field, the spectral function $A(\omega)$ and the e-ph SE $-\Im[\Sigma^{\text{R}}_{\text{e-ph}}(\omega)]$ are displayed in Fig.~\ref{fig:EQ_PH_SF_PH_SE}. $A(\omega)$ features Hubbard bands centered at $\omega_{\text{HB},\pm} = \pm 3.93t^{\ast}$ and no \emph{quasi-particle} peak at $\omega = 0$. Similarly, $-\Im[\Sigma^{\text{R}}_{\text{e-ph}}(\omega)]$ shows a gap with bands centered at $\omega_{\text{SE},\pm} = \pm 4.17t^{\ast}$. The energy shift between the e-ph  SE and Hubbard bands amounts to about twice the phonon frequency $| \omega_{\text{HB},\pm} - \omega_{\text{SE},\pm} | \ = 0.24t^{\ast} \approx 2\omega_{\text{ph}}$. 
 
\begin{figure}[t]
\includegraphics[width=\linewidth]{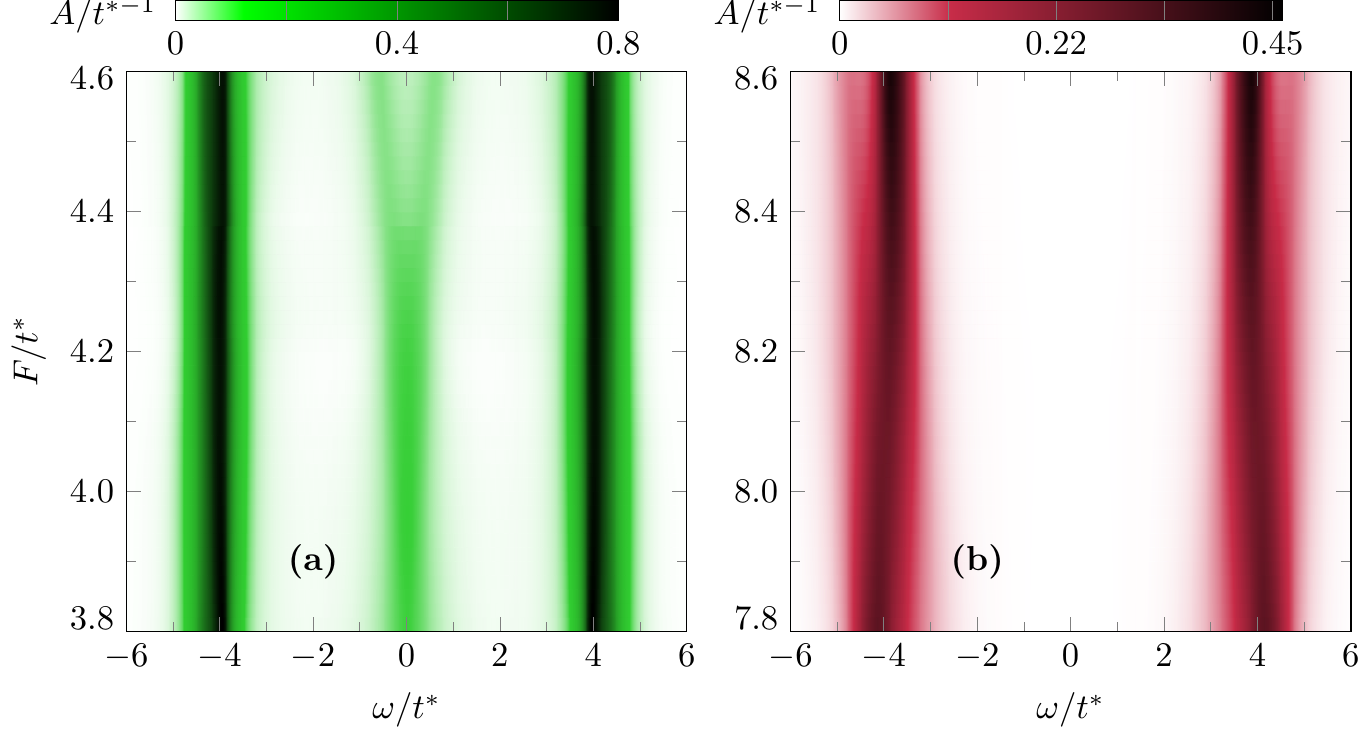}
\caption{(Color online) Color map of the spectral function $A(\omega)/t^{\ast-1}$ in the accessible regions (a) $F/t^{\ast}\in \{ 3.80,3.82 \dots, 4.60 \}$ and (b) $F/t^{\ast}\in \{ 7.80,7.82 \dots, 8.60 \}$. Default parameters are specified at the beginning of section~\ref{sec:results}.}
\label{fig:spectrum_PH_maxima}
\end{figure}

As evidenced by Fig.~\ref{fig:EQ_PH_SF_PH_SE}, the e-ph SE $-\Im(\Sigma^{\text{R}}_{\text{e-ph}})$ is gapped and thus does not provide a dissipation channel over the whole frequency interval. This limits a stable steady-state solution to values of $F$ close to the two main resonances $F/t^{\ast} \approx U/2$  and  $U$, the spectra of which are shown in Fig.~\ref{fig:spectrum_PH_maxima}.
As in the previous setup, the WS side-bands are present as can be seen in Fig.~\ref{fig:spectrum_PH_maxima}(a): their merging at $F \approx U/2 \approx \omega_{\text{HB},\pm}$ creates the necessary in-gap spectral weight that allows the occupation of the UHB as discussed in Sec.~\ref{sec:e+ph+spectra}. 

\begin{figure}[b]
\includegraphics[width=\linewidth]{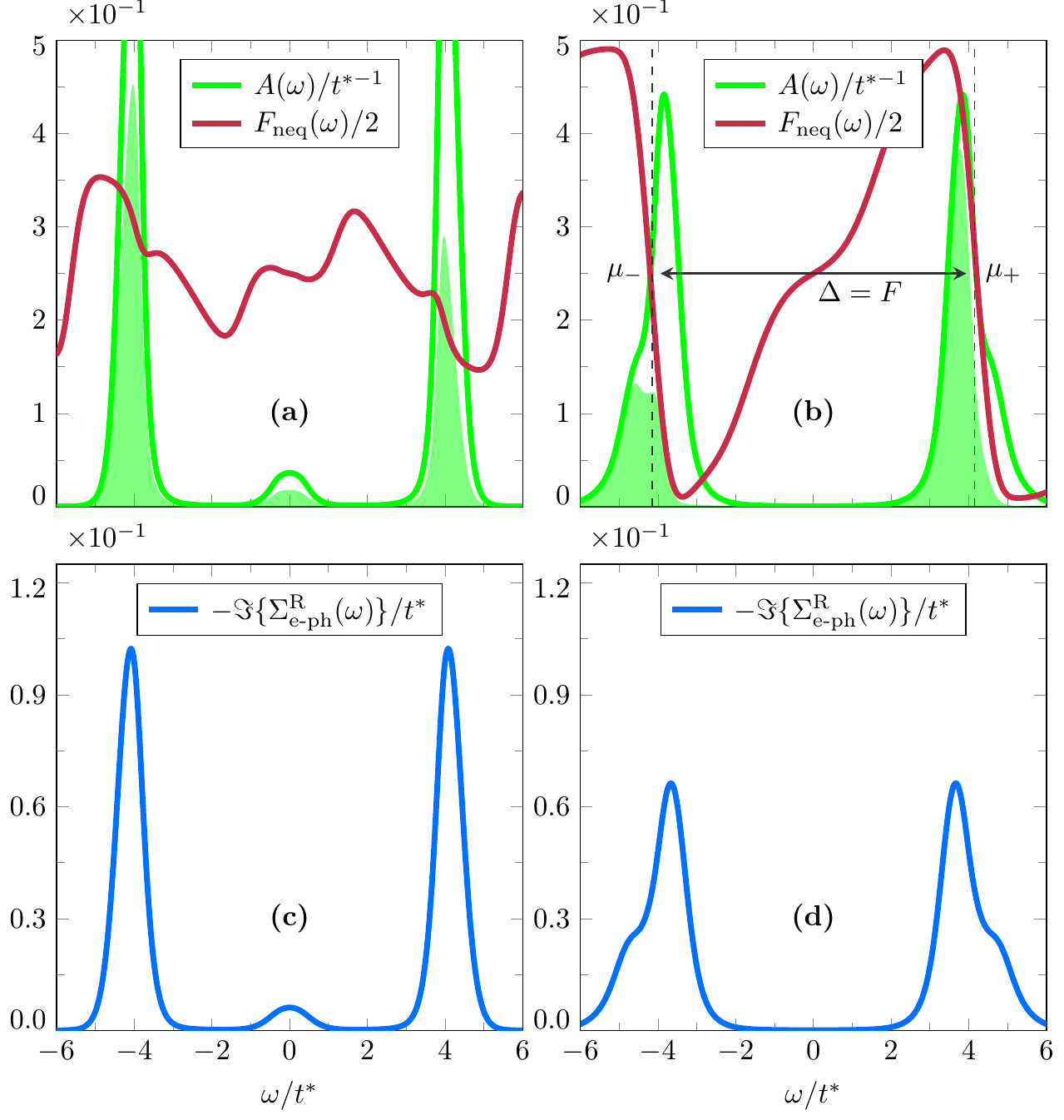}
\caption{(Color online) [(a),(c)] \resub{Spectral function} $A(\omega)/t^{\ast-1}$, \resub{spectral occupation $N_{\text{e}}(\omega)/t^{\ast-1}$} (shaded area), \resub{non-equilibrium distribution function} $F_{\text{neq}}(\omega)$ and \resub{e-ph SE} $-\Im[\Sigma^{\text{R}}_{\text{e-ph}}(\omega)]/t^{\ast}$ at $F/t^{\ast}=4.20$. [(b),(d)] Same quantities for $F/t^{\ast}=8.42$. The arrow highlights the energy splitting $\Delta = F$ that introduces effective chemical potentials for 
the Hubbard bands. Default parameters are specified at the beginning of section~\ref{sec:results}.}
\label{fig:PH_SFs_maxima}
\end{figure}

\begin{figure}[t]
\includegraphics[width=\linewidth]{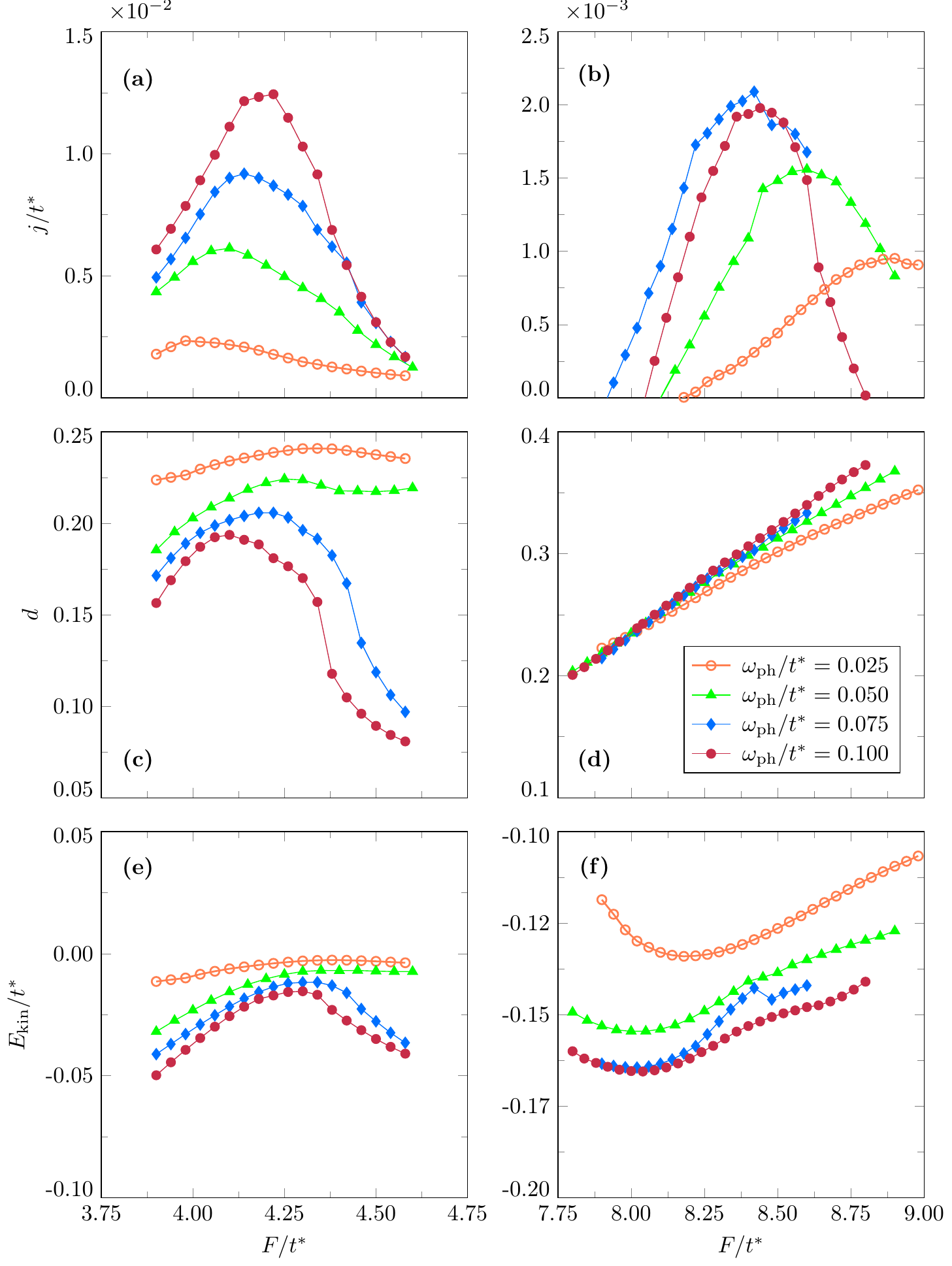}
\caption{(Color online) Electric field dependence of the current [(a),(b)], double occupation [(c),(d)] and kinetic energy [(e),(f)] near the resonances $F/t^{\ast}\approx U/2$ and $F/t^{\ast}\approx U$, for $\omega_{\text{ph}}/t^{\ast}=\lbrace 0.025, 0.050, 0.075, 0.100 \rbrace$. The default value of $\Gamma_{\text{ph}}/t^{\ast}$ is specified at the beginning of section~\ref{sec:results}.}
\label{fig:phonon_observables} 
\end{figure}

Fig.~\ref{fig:PH_SFs_maxima}(a) shows a cut of the spectrum displayed in Fig.~\ref{fig:spectrum_PH_maxima}(a) at $F=4.2t^{\ast}\approx \omega_{\text{HB},+} + 2\omega_{\text{ph}} \approx \omega_{\text{SE},+}$, where the current (Fig.~\ref{fig:phonon_observables}(a)) reaches its first maximum. We note that the spectral weight and amplitude of the states around $\omega=0$ is roughly the same as in Fig.~\ref{fig:SF+Fill_gamma0065}(b) while the Hubbard bands are narrower, confirming that the fermionic bath provides an overall broadening but has no influence on the transport properties mediated by the in-gap sub-bands. The corresponding imaginary part of the \emph{retarded} e-ph SE shown in Fig.~\ref{fig:PH_SFs_maxima}(c) also exhibits a dissipative region around $\omega=0$, which effectively halves the gap, and facilitates transitions from LHB to UHB. 

For field strengths $F\gtrsim4.6t^{\ast}$ the dissipation by phonons is strongly suppressed: this can be explained by the reduction of the in-gap spectral weight at $\omega=0$ due to the parting WS side-bands as the field is further increased, which is visible in Fig.~\ref{fig:spectrum_PH_maxima}(a). The net effect is that the total spectrum of the in-gap states is no longer sufficient to allow particle decay across the gap~\footnote{\resub{Note that as the overall magnitude of the in-gap spectrum diminishes, so does $\Im [\Sigma^{\text{R}}_{\text{e-ph}}]$. Relaxations of high energy carriers through the gap are only possible if the spectral weight of the e-ph SE is not too small, see the discussion in Sec.~\ref{sec:e+ph+spectra} and Fig.~\ref{fig:SF+Fill_gamma0065}(b).}}. 

On the other hand, at field strengths crossing the gap, e.g. $F=2\omega_{\text{HB},+} \approx 7.86t^{\ast}$, phonons are expected to dissipate again. In particular, we find that the current increases and reaches its second maximum at $F=8.42t^{\ast} \approx 2\left( \omega_{\text{HB},+} + 2\omega_{\text{ph}} \right) \approx 2\omega_{\text{SE},+}$ (Fig. \ref{fig:phonon_observables}(b)). The corresponding spectral function and e-ph SE profiles are shown in Fig.~\ref{fig:PH_SFs_maxima}(b) and (d). Here the Hubbard bands exhibit a population-inversion, as evidenced by the spectral occupation $N_{\text{e}}(\omega)$ and the non-equilibrium distribution function $F_{\text{neq}}(\omega)$. The latter can be approximately described in terms of two effective chemical potentials $\mu_{\pm}\approx \omega_{\text{SE},\pm}$ separated by the resonant field, with a larger distribution in the UHB. 
The steady-state is thus characterized by fewer empty states available for transport, resulting in a current, which is almost one order of magnitude smaller than the typical values obtained with the coupling to a fermionic bath. The dissipation mechanism is less effective for phonons alone than in the presence of a fermionic bath. Expanding the non-equilibrium distribution to linear order around the effective chemical potentials as $F_{\text{neq}}(\omega\sim \mu_{\pm})-1/2\approx -(\beta_{\text{eff}}/4)(\omega-\mu_{\pm})$, one can extract two equal effective temperatures $\beta_{\text{eff}}\approx 9/t^{\ast}$~\footnote{Due to particle-hole symmetry, the two temperatures coincide.}, while phonons are kept at $\beta=20/t^{\ast}$.

\begin{figure}[t]
\includegraphics[width=\linewidth]{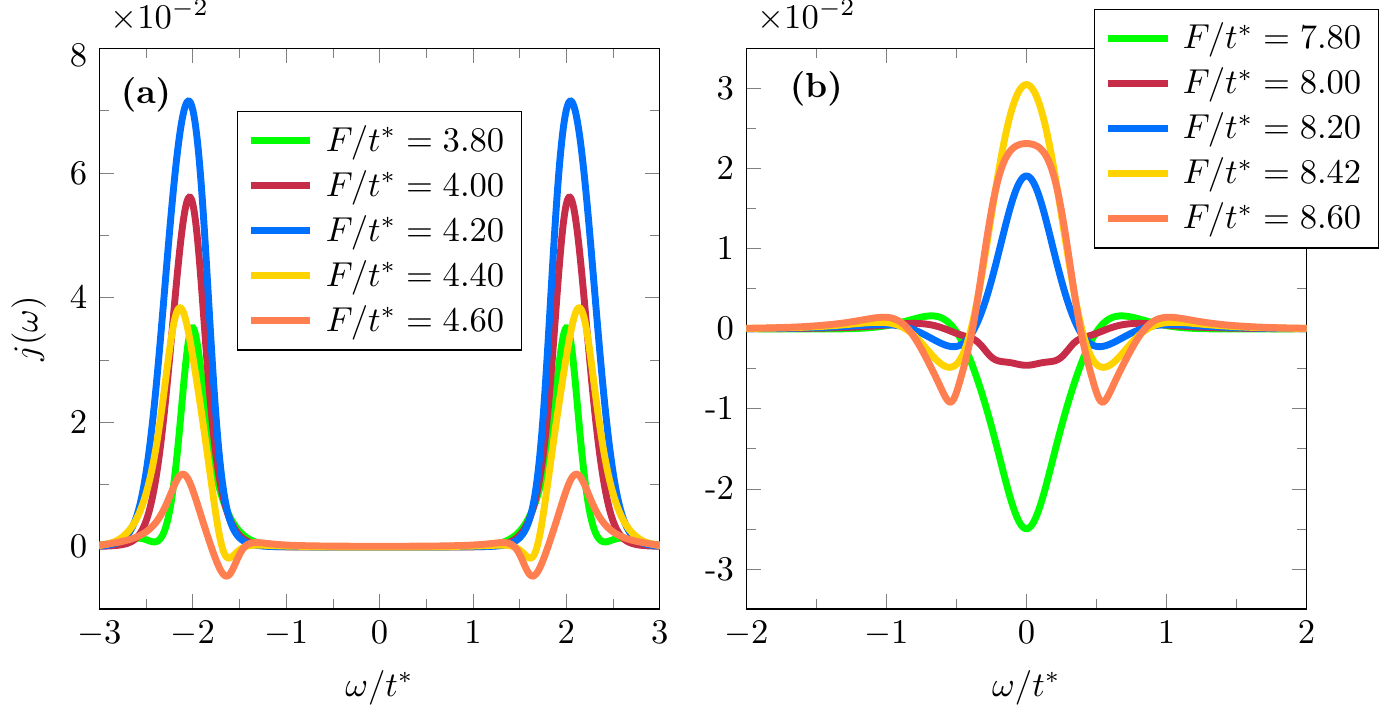}
\caption{(Color online) Frequency resolved current $j(\omega)$ near the resonances at (a) $F/t^{\ast}\approx U/2$ and (b) $F/t^{\ast}\approx U$. Default parameters are specified at the beginning of section~\ref{sec:results}.}
\label{fig:Js_ph_maxima}
\end{figure}

\subsubsection{Current, energy and double occupancy}\label{sec:PH_observables}
As pointed out in Sec.~\ref{sec:PH_SF_Filling}, phonons alone dissipate around the two main resonances $F\approx U/2$ and $U$. Our results show that the actual peaks in the current are shifted by the phonon frequency and occur at $F=4.2\approx \omega_{\text{HB},+}+2\omega_{\text{ph}}$ and $F=8.42\approx 2(\omega_{\text{HB},+}+2\omega_{\text{ph}})$ as shown in Fig.~\ref{fig:phonon_observables}(a) and (b). This implies that the energy range over which phonons affect the system is closely related to their characteristic frequency $\omega_{\text{ph}}$, see also Fig.~\ref{fig:EQ_PH_SF_PH_SE}. 

On the other hand, the slight imbalance in the current at $F\approx U/2$ and $F\approx U$ in favor of the former resonance noted for Fig.~\ref{fig:Observables_bothbaths}(a) is even stronger in the case of phonon-only dissipation, as evidenced by the suppression of the second resonance peak in Fig.~\ref{fig:phonon_observables}(b) (notice the scale).

\begin{figure}[b]
\includegraphics[width=0.9\linewidth]{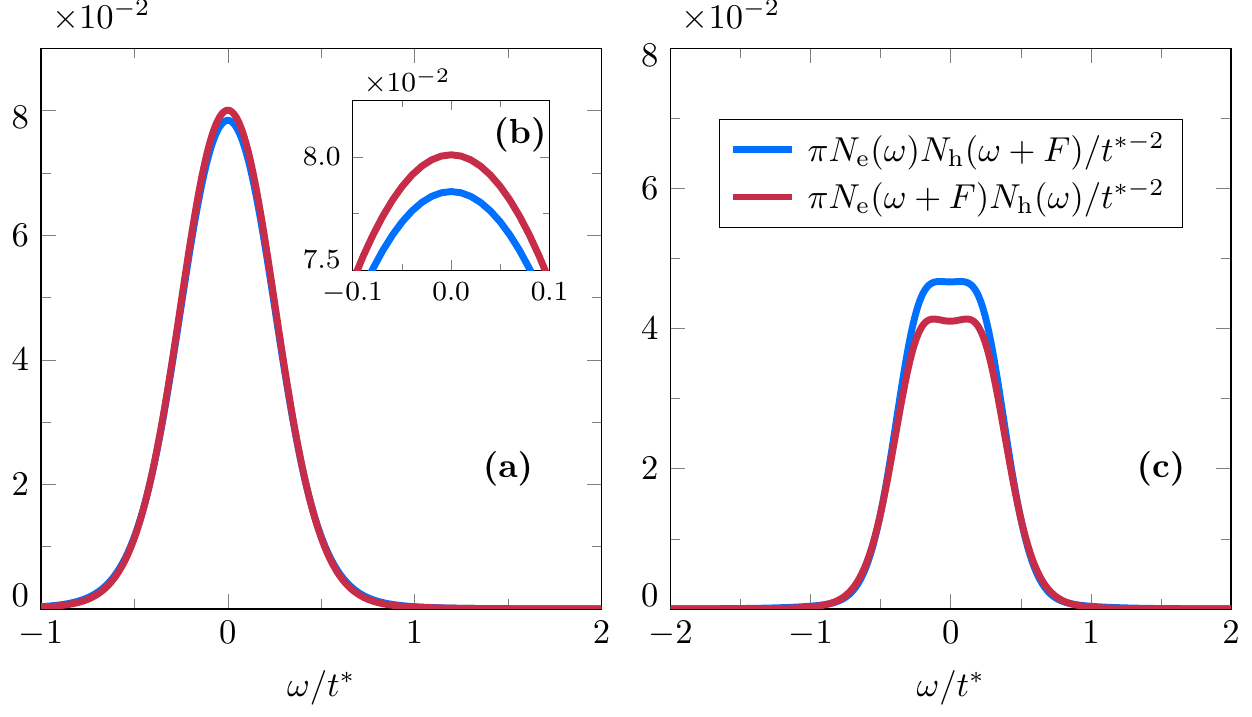}
\caption{(Color online) Forward (blue) and backward (red) flow contributions to the frequency-dependent tunneling current~\eqref{eq:jtun}. [(a),(b)] For $F=7.8t^{\ast}$, the backflow overcomes the forwards flow and leads to a negative current. (c) At $F=8.42t^{\ast}$, the forwards flow dominates. Default parameters are specified at the beginning of section~\ref{sec:results}.}
\label{fig:PH_SFs_tunnel}
\end{figure}

\begin{figure*}[t]
\includegraphics[width=0.7\linewidth]{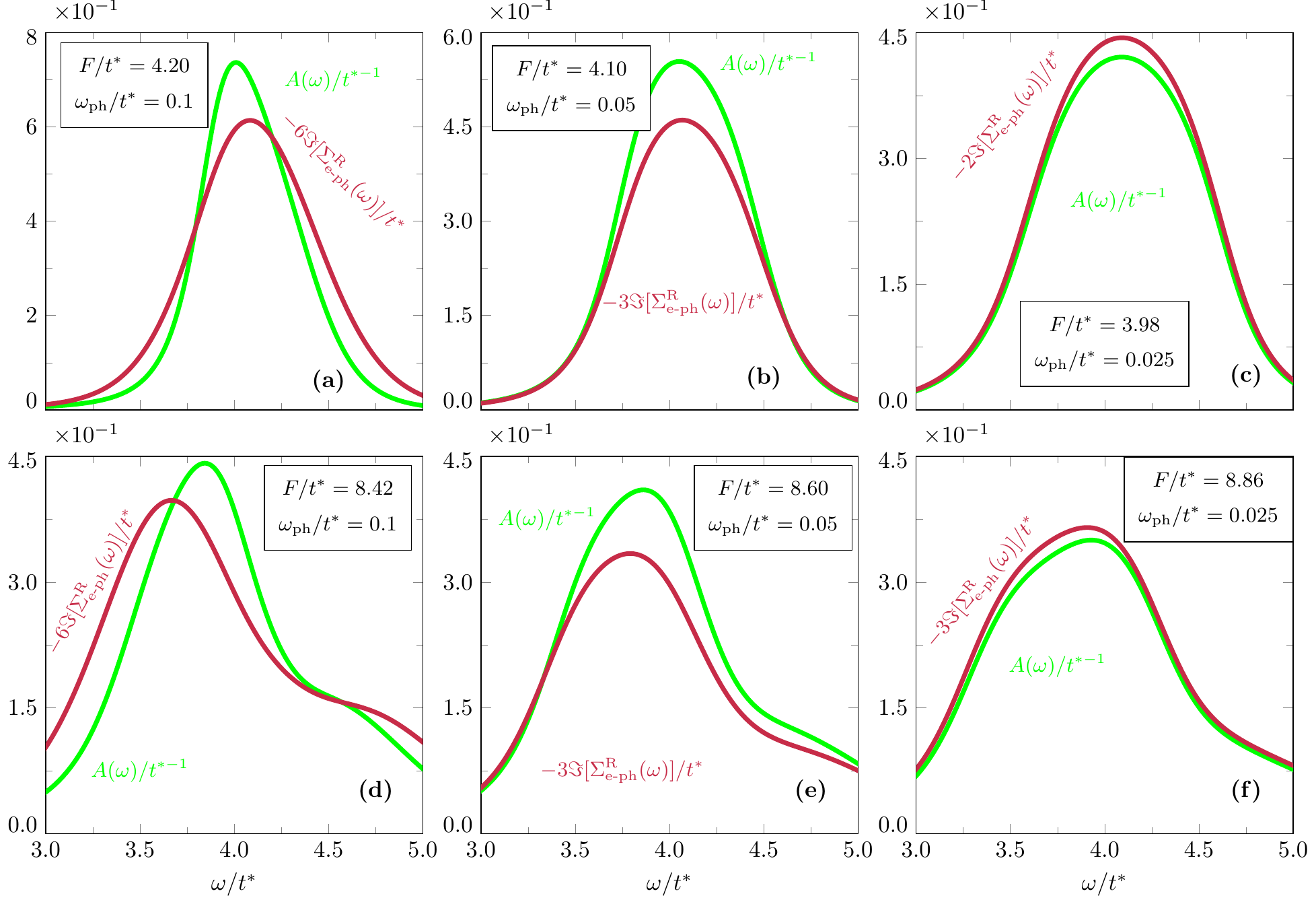}
\caption{\resub{[(a),(d)] Spectral function $A(\omega)/t^{\ast-1}$ and e-ph SE $-\Im[\Sigma^{\text{R}}_{\text{e-ph}}(\omega)]/t^{\ast}$ displayed for $\omega_{\text{ph}}/t^{\ast}=0.1$ at field strengths which correspond to the current maxima in Figs.~\ref{fig:phonon_observables}(a)-(b). Same quantities for [(b),(e)] $\omega_{\text{ph}}/t^{\ast}=0.05$ and [(c),(f)] $\omega_{\text{ph}}/t^{\ast}=0.025$. Default parameters are specified at the beginning of section~\ref{sec:results}.}}
\label{fig:shifting_Jpeaks_ph_only}
\end{figure*}

For a deeper understanding of the current characteristics, we now focus on the frequency resolved current $j(\omega)$ for field strengths near the resonances mentioned above. Away from the resonant value $F= 4.2 t^{\ast}\approx \omega_{\text{HB},+}+2\omega_{\text{ph}}$, the frequency resolved current $j(\omega)$ starts to take negative values, see Fig.~\ref{fig:Js_ph_maxima}(a). This tendency increases at higher field strengths $F\approx7.8t^{\ast}$, Fig.~\ref{fig:Js_ph_maxima}(b). 

Once again the tunnelling formula~\eqref{eq:jtun} provides more insight which is illustrated in Fig.~\ref{fig:PH_SFs_tunnel}: when the field is close to the resonance $F\approx U$, particles start to be resonantly promoted from the LHB to the UHB. Lacking the fermionic bath, the accumulation of high energy carriers within the UHB is due to the backflow $N_{\text{e}}(\omega+F)N_{\text{h}}(\omega)$ which can overcome the forwards flow $N_{\text{e}}(\omega)N_{\text{h}}(\omega+F)$ as shown in Fig.~\ref{fig:PH_SFs_tunnel}(a), see also Ref.~\cite{le.pa.14}. Notably as the field approaches $F=8.42t^{\ast} \approx 2(\omega_{\text{HB},+}+2\omega_{\text{ph}})$, the contribution $N_{\text{e}}(\omega)N_{\text{h}}(\omega+F)$ is sufficiently large to ensure a positive current as seen in Figs.~\ref{fig:phonon_observables}(b), \ref{fig:Js_ph_maxima}(b) and \ref{fig:PH_SFs_tunnel}(b) despite the large number of high energy occupied states in the UHB visible in Fig.~\ref{fig:PH_SFs_maxima}(b).
 
The double occupation per site and the kinetic energy around the two main resonances are shown in \resub{Figs.}~\ref{fig:phonon_observables}(c)-(f). At $F\approx U/2$ (\resub{Figs.}~\ref{fig:phonon_observables}(c) and (e)), both $d$ and $E_{\text{kin}}$ follow the same bahavior of $j$. Specifically, they display an ascending phase followed by a downturn. Their maxima are slightly shifted from each other, in agreement with the discussion in Sec.~\ref{sec:observables_bothbaths}. More interesting is the behavior near $F\approx U$ (\resub{Figs.}~\ref{fig:phonon_observables}(d) and (f)). There, the double occupation $d$ increases linearly as a function of $F$ while the kinetic energy $E_{\text{kin}}$ first reaches a minimum and then starts to increase as well. The linear increase of the double occupation signals resonant driving and the lack of fermion baths which no longer drain the UHB and inject the particles back in the LHB. For the kinetic energy, the depletion of particles in the LHB implies that less energy can be gained by accelerating particles. In summary, more energy is used to overcome the interaction energy than to increase the kinetic energy. We also note, that while in the case of fermionic baths, $d$ never exceeded its \emph{uncorrelated} value $d_{\text{uncorr}}=0.25$, in case of only phonons, $d$ can increase beyond this value, confirming the accumulation of high-energy doublons seen in Fig.~\ref{fig:PH_SFs_maxima}(b). 

We now investigate how the current depends on the characteristic phonon frequency $\omega_{\text{ph}}$. The first resonance sensitively depends on $\omega_{\text{ph}}$ as can be seen in Fig.~\ref{fig:phonon_observables}(a). Its maximum is shifted to $F\approx \omega_{\text{HB},+} + 2\omega_{\text{ph}}$, as already noted in Sec.~\ref{sec:PH_SF_Filling}. Also, $j$ grows linearly upon increasing $\omega_{\text{ph}}$ at this field strength. Considering the current at $F\approx U/2$, the difference between our setup with electronic baths and phonons at $\gamma=0.065$ discussed in Sec.~\ref{sec:WBL_plus_ph} and the one presented in Ref.~\cite{mu.we.18} amounts to the value obtained in the phonon-only setup shown in Fig.~\ref{fig:phonon_observables}(a) for $\omega_{\text{ph}}/t^{\ast}=0.1$.

\resub{The peak at $F\approx U$ displayed in Fig.~\ref{fig:phonon_observables}(b) behaves partly different. While the height of the maximum also increases with $\omega_{\text{ph}}$, the resonance moves to lower electric fields as the cutoff frequency is increased. This behavior can be explained using Fig.~\ref{fig:shifting_Jpeaks_ph_only}, which shows the spectral function and e-ph SE for the peak electric fields belonging to the different cutoff frequencies of Fig.~\ref{fig:phonon_observables}. First notice that in the first row~(a)-(c) belonging to $F\approx U/2$, the e-ph SE shares the same support as the spectral function. In contrast to this, in the second row~(d)-(f) belonging to $F\approx U$, a larger cutoff frequency $\omega_{\text{ph}}/t^{\ast}=0.1$ shifts the e-ph SE closer to the gap. This may be due to the increasing $\omega_{\text{ph}}$ broadening the phonon spectral function $A_{\text{ph}}(\omega)$ which thus also broadens the e-ph SE. The broader e-ph SE provides the necessary spectral weight for particle-relaxation across the gap, in agreement with the discussion in Sec.~\ref{sec:PH_SF_Filling}.

In summary the resonance in the current at $F\approx U/2$ is determined by the peak positions of the LHB and UHB which appear to be pinned by the peaks of the e-ph SE as discussed for Fig.~\ref{fig:EQ_PH_SF_PH_SE}. Here, the peak positions are decisive, since the particles move via spectral weight located around $\omega=0$, see Fig.~\ref{fig:PH_SFs_maxima}(a) and (c). For the resonance at $F\approx U$, the spectral weight at both gap edges and thus the extent of the spectral function into the gap is relevant. Here the broadening effect of $\omega_{\text{ph}}$ seems to determine the resonant electric field.}

\section{Conclusions}\label{sec:conclusions}
We investigated the single-particle spectrum and the current characteristics of a Mott insulating system under the action of a strong dc electric field. In this work we highlighted the effects of acoustic phonons as dominant mechanism for providing dissipation thus enabling a stationary dc current, in contrast to fermionic baths adopted in previous work~\cite{mu.we.18,am.we.12,aron.12,li.ar.15}.

Starting with the case of both phonons and weakly coupled fermionic baths, we observe the occurrence of in-gap states and transport resonances associated with Wannier-Stark side-bands similar to previous work~\cite{mu.we.18,aron.12}. In case of phonons as the only source of dissipation, we find them to be much less effective, meaning that a steady-state solution is achieved only for electric fields close to the resonances. In these regions, we find a direct correlation between the resonances' location and the phonon characteristic frequency. In addition, we observe a strong population shift to higher bands around the resonances and in particular a population inversion at large fields which can be described in terms of two separate chemical potentials for the LHB and the UHB. These effects are not observed for the mixed fermion-phonon dissipation mechanism due to the constant drain and emission of carriers by the fermionic baths. When phonons are the {\em only} dissipation mechanism, the current in the metallic phase is almost one order of magnitude smaller than the typical values obtained by coupling to the fermionic bath. 

\resub{Employing the so-called {\em self-consistent} Migdal approximation which allows phonons to heat up via back-action from electrons is expected to display an even less effective dissipation characterized by a further reduction of the current due to hot-phonon effects at the onset of the metallic phase. However, acoustic phonons could still be able to dissipate to some extent because of their finite, albeit small, bandwidth. This aspect will be investigated in a future work.}

As pointed out in previous works~\cite{ec.we.13,mu.we.18,ap.st.12}, 
experimentally the quasi-stationary situation considered here cannot be induced by a regular dc field, since it would have to be far too strong and destroy the material. \resub{Following the reasoning in Ref.~\cite{mu.we.18}, high electric field amplitudes $F\sim\unit[1]{V/\AA}$ corresponding to our units are available in the THz regime from few cycle pulses which are used for high harmonic generation from solids~\cite{gh.di.11,gh.re.19}. They argue that below driving frequencies of the order a few hundreds THz some of the spectral features could be safely described by a {\em quasi-dc} setup. To our knowledge, unfortunately such WS excitations have not been observed so far.}

\begin{acknowledgments}
We thank A. Picano for fruitful discussions. This research was funded in part by the Austrian Science Fund (Grant No. P 33165-N) and by NaWi Graz. The computational results presented have been obtained using the Vienna Scientific Cluster and the D-Cluster Graz.
\end{acknowledgments}

\appendix 

\section{The $d$-dimensional current and kinetic energy}\label{sec:ss_j_Ekin_derivation}
We start from the \emph{equal-time} GF $G^{<}_{\vec{k}}(t,t)\equiv G^{<}_{\vec{k}}(t_{\text{av}},t_{\text{rel}}=0)$, which follows from Eq.~\eqref{eq:WignerGF} as
\begin{equation}\label{eq:equal_timeGF}
G^{<}_{\vec{k}}(t_{\text{av}},t_{\text{rel}}=0) = \sum_{l^{\prime}=-\infty}^{+\infty} \int_{-\infty}^{+\infty} \frac{\dd\omega}{2\pi} \ G^{<}_{l^{\prime}}(\omega,\vec{k}) e^{-il\Omega t_{\text{av}}}.
\end{equation}
Following Ref.~\cite{ts.ok.08}, the $l$-th Wigner component of Eq.~\eqref{eq:time_current} is given by $j_{l}(\Omega) = \int_{-\frac{\tau}{2}}^{\frac{\tau}{2}} \frac{\dd t_{\text{av}}}{\tau} \ e^{\ii l\Omega t_{\text{av}}} j(t_{\text{av}})$ which leads combined with Eq.~\eqref{eq:equal_timeGF} to
\begin{equation}\label{eq:dim_dep_Wig_current}
	j_{l}(\omega) = 2i \sum_{l^{\prime}=-\infty}^{+\infty} \sum_{\vec{k}} \ [\vec{e}_{0} \cdot \nabla_{\vec{k}}(\varepsilon_{\vec{k}})_{l-l^{\prime}} ] G^{<}_{l^{\prime}}(\omega,\vec{k}).
\end{equation}
The $d$-dimensional $\vec{k}$-space derivative in Eq.~\eqref{eq:dim_dep_Wig_current} reads 
\begin{equation}\label{eq:dDim_curr_proj}
\begin{split}
	\vec{e}_{0} \cdot \nabla_{\vec{k}} (\epsilon_{\vec{k}})_{l-l^{\prime}} & = \sum_{n=1}^{d} \frac{\partial}{\partial k_n} (\epsilon_{\vec{k}})_{l-l^{\prime}} \\
	& = \sum_{n=1}^{d} \left[ \frac{\partial \epsilon}{\partial k_n} \frac{\partial (\epsilon_{\vec{k}})_{l-l^{\prime}}}{\partial \epsilon} + \frac{\partial \overline{\epsilon}}{\partial k_n} \frac{\partial (\epsilon_{\vec{k}})_{l-l^{\prime}}}{\partial \overline{\epsilon}} \right],
\end{split}
\end{equation}
where by means of the Floquet dispersion relation in Eq.~\eqref{eq:Floquet_disp}, one finds
\begin{equation}\label{eq:der_disp_k_dc}
\begin{split}
	\frac{\partial (\epsilon_{\vec{k}})_{l-l^{\prime}}}{\partial \epsilon} & = \frac{1}{2} \left[ \delta_{l-l^{\prime},1} + \delta_{l-l^{\prime},-1} \right], \\ 
	\frac{\partial (\epsilon_{\vec{k}})_{l-l^{\prime}}}{\partial \overline{\epsilon}} & = \frac{i}{2} \left[ \delta_{l-l^{\prime},1} - \delta_{l-l^{\prime},-1} \right]. \\
\end{split}
\end{equation}
Since Eq.~\eqref{eq:der_disp_k_dc} does not depend on the index $n$, one can make use of the explicit form of $\epsilon$ and $\overline{\epsilon}$ in Eq.~\eqref{eq:d-dim_crystal_dep} and write  
\begin{equation}\label{eq:der_eps_epsbar_dc}
\begin{split}
\sum_{n=1}^{d} \frac{\partial \epsilon}{\partial k_n} & = -\overline{\epsilon}, \\ 
\sum_{n=1}^{d} \frac{\partial \overline{\epsilon}}{\partial k_n} & = \epsilon. \\
\end{split}
\end{equation}
By inserting Eqs.~\eqref{eq:der_disp_k_dc} and \eqref{eq:der_eps_epsbar_dc} into Eq.~\eqref{eq:dDim_curr_proj} while recalling that the sum over $\vec{k}$ can be performed over $(\epsilon,\overline{\epsilon})$, the current integrand is 
\begin{equation}\label{eq:Curr_integrand_eps_epsbar_int}
\begin{split}
j_{l}(\omega) & = \int \dd\epsilon \int \dd\overline{\epsilon} \ \rho(\epsilon,\overline{\epsilon}) \times \\
&\times \left[ - \left( \epsilon + i \overline{\epsilon} \right) G^{<}_{l-1}(\omega,\epsilon,\overline{\epsilon}) + \left( \epsilon -i \overline{\epsilon} \right) G^{<}_{l+1}(\omega,\epsilon,\overline{\epsilon}) \right],  
\end{split}
\end{equation}
which can be recast as in Eq.~\eqref{eq:Current_integrald_L} exploiting the symmetry property $\left(\mat{G}^{<}\right)^{\dagger} = -\mat{G}^{<}$ and the Floquet-to-Wigner mapping provided in Eq.~\eqref{eq:Wig2Fl}. 

We now summarize the main steps to derive the kinetic energy integrand in Eq.~\eqref{eq:Energy_integrald_L}. The starting point is the analog of Eq.~\eqref{eq:dim_dep_Wig_current}, i.e. 
\begin{equation}\label{eq:dim_dep_Wig_energy}
E_{\text{kin},l}(\omega) = -2i \sum_{l^{\prime}=-\infty}^{+\infty} \sum_{\vec{k}} \ (\varepsilon_{\vec{k}})_{l-l^{\prime}} G^{<}_{l^{\prime}}(\omega,\vec{k})
\end{equation}
expressing the $l$-th Wigner component of the kinetic energy integrand. As in the previous case, by inserting the Floquet dispersion relation~\eqref{eq:Floquet_disp} in Eq.~\eqref{eq:dim_dep_Wig_energy} and making use of the JDOS, we get 
\begin{equation}\label{eq:Nrg_integrand_eps_epsbar_int}
\begin{split}
E_{\text{kin},l}(\omega) & = \int \dd\epsilon \int \dd\overline{\epsilon} \ \rho(\epsilon,\overline{\epsilon}) \times \\
&\times \left[ (\overline{\epsilon} - i \epsilon) G^{<}_{l-1}(\omega,\epsilon,\overline{\epsilon}) - (\overline{\epsilon} + i \epsilon) G^{<}_{l+1}(\omega,\epsilon,\overline{\epsilon}) \right],
\end{split}
\end{equation}
which is equivalent to Eq.~\eqref{eq:Energy_integrald_L} as it can be proven by using the same properties mentioned above. 

\section{Real-time Keldysh components of the electron-phonon SE}\label{sec:real-time_eph_se}

\begin{figure}
\includegraphics[width=\linewidth]{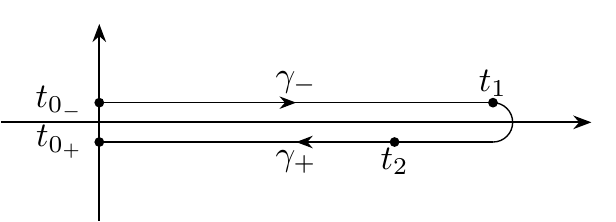}
\caption{Keldysh contour $C_{\kappa}=\gamma_{-} \cup \gamma_{+}$ for real-time arguments.}
\label{fig:keldysh_contour}
\end{figure}

The Keldysh components of the e-ph SE in Eq.~\eqref{eq:backbone_e-ph_SE} are explicitly derived by means of the \emph{Langreth rules} \cite{st.va.13} and read
\begin{align}\label{eq:TD_e-ph_SE}
\begin{split}
\Sigma^{\text{R}}_{\text{e-ph}}(t,t^{\prime})& = ig^{2} \left[ G^{\text{R}}(t,t^{\prime})\resub{D^{>}_{\text{ph}}}(t,t^{\prime}) + G^{<}(t,t^{\prime})\resub{D^{\text{R}}_{\text{ph}}}(t,t^{\prime}) \right], \\
\Sigma^{\text{K}}_{\text{e-ph}}(t,t^{\prime}) & = ig^{2} \left[ G^{\text{K}}(t,t^{\prime})\resub{D^{>}_{\text{ph}}}(t,t^{\prime}) \ + \right. \\ 
& \left. + \ G^{<}(t,t^{\prime}) \left(\resub{D^{>}_{\text{ph}}}(t,t^{\prime})-\resub{D^{<}_{\text{ph}}}(t,t^{\prime})\right) \right], \\
\end{split}
\end{align}
where $t,t^{\prime}$ lie on the Keldysh contour $C_{\kappa}$ shown in Fig.~\ref{fig:keldysh_contour}.

\section{Phonon spectral function}\label{sec:phonon_spectrum}
The dispersion relation of acoustic phonon branches can be approximated~\cite{gr.pa.14} by
\begin{equation}\label{eq:linear_PH_disp_rel}
\omega(\vec{q}) \approx v_{\text{s}} |\vec{q}|
\end{equation}
for reciprocal lattice vectors $\vec{q}$ satisfying
 $|\vec{q}| a \ll 1$, where $v_{\text{s}}$ is the characteristic speed of sound of the considered lattice, and $a$ the lattice spacing. 
In $d$ dimensions the phonon modes' density of states is 
\begin{equation}\label{eq:modes_dDOS}
D_{d}(\omega)=\sum_{\vec{q}} \delta[\omega-\omega(\vec{q})]=\left( \frac{a}{2\pi}\right)^{d} \int \dd^{d}q \ \delta[\omega-\omega(\vec{q})].
\end{equation}
By taking into account $d=\{ 1,2,3 \}$ and making use of polar coordinates we find
\begin{align}\label{eq:123dDOS_general}
\begin{split}
D_{1}(\omega)=& \ \frac{a}{2\pi} \ \int_{0}^{+\infty} \dd|\vec{q}| \ \delta(\omega-v_{\text{s}}|\vec{q}|), \\
D_{2}(\omega)=& \left( \frac{a}{2\pi}\right)^{2} 2\pi \int_{0}^{+\infty} \dd|\vec{q}| \ |\vec{q}| \delta(\omega-v_{\text{s}}|\vec{q}|), \\
D_{3}(\omega)=& \left( \frac{a}{2\pi}\right)^{3} 4\pi \int_{0}^{+\infty} \dd|\vec{q}| \ |\vec{q}|^{2} \delta(\omega-v_{\text{s}}|\vec{q}|). \\
\end{split}
\end{align}
Recalling that $\delta(\omega-v_{\text{s}}|\vec{q}|) = \frac{1}{v_{\text{s}}}\delta(|\vec{q}|-\omega/v_{\text{s}})$, we then get  
\begin{align}\label{eq:123dDOS}
\begin{split}
D_{1}(\omega)=& \ \frac{a}{2\pi} \frac{1}{2v_{\text{s}}},  \\
D_{2}(\omega)=& \frac{a^{2}}{\pi} \frac{\omega}{4v^{2}_s}, \\
D_{3}(\omega)=& \frac{a^{3}}{2\pi^{2}} \frac{\omega^{2}}{2v^{3}_s}. \\
\end{split}
\end{align}
A rough estimate for $v_{\text{s}}$ can be obtained as $v_{\text{s}}\approx\omega_{\text{ph}}a$. Here $\omega_{\text{ph}}$ is the characteristic frequency for which the phonon modes dispersion relation is still linear, i.e., $\omega_{\text{ph}}=\omega(\vec{q}_{\text{max}})$. Accordingly $\vec{q}_{\text{max}}$ is the maximal value the reciprocal lattice vector can take for Eq.~\eqref{eq:linear_PH_disp_rel} to be fulfilled. 

\bibliographystyle{prsty}
\bibliography{references_database,my_refs}

\end{document}